\documentclass[1p,12pt]{elsarticle}
\usepackage{graphicx}
\usepackage{epstopdf}
\usepackage{amssymb} 
\usepackage{amsmath} 
\usepackage{color} 
\usepackage{axodraw}

\usepackage{dcolumn}

\newcommand{\ifm}[1]{\relax\ifmmode #1\else $#1$5\fi}
 
\newcommand{\beq}{\begin{equation}}
\newcommand{\eeq}{\end{equation}}
\newcommand{\beqn}{\begin{eqnarray}}
\newcommand{\eeqn}{\end{eqnarray}}
\newcommand{\bi}{\begin{itemize}}
\newcommand{\ei}{\end{itemize}}
\newcommand{\bd}{\begin{description}}
\newcommand{\ed}{\end{description}}
\newcommand{\bHuge}{\begin{Huge}}
\newcommand{\bhuge}{\begin{huge}}
\newcommand{\bLARGE}{\begin{LARGE}}
\newcommand{\bLarge}{\begin{Large}}
\newcommand{\blarge}{\begin{large}}
\newcommand{\eHuge}{\end{Huge}}
\newcommand{\ehuge}{\end{huge}}
\newcommand{\eLARGE}{\end{LARGE}}
\newcommand{\eLarge}{\end{Large}}
\newcommand{\elarge}{\end{large}}

\def \gtsim    {\relax\ifmmode{\mathrel{\mathpalette\oversim >}}
                  \else{$\mathrel{\mathpalette\oversim >}$}\fi}
\def \ltsim    {\relax\ifmmode{\mathrel{\mathpalette\oversim <}}
                  \else{$\mathrel{\mathpalette\oversim <}$}\fi}
\def\oversim#1#2{\lower4pt\vbox{\baselineskip0pt \lineskip1.5pt
            \ialign{$\mathsurround=0pt#1\hfil##\hfil$\crcr#2\crcr\sim\crcr}}}
\def \dk       {\relax\ifmmode{\rightarrow}\else{$\rightarrow$}4\fi}
\def \to       {\relax\ifmmode{\rightarrow}\else{$\rightarrow$}4\fi}
\def \Dk    {\relax\ifmmode{\Rightarrow}\else{$\Rightarrow$}\fi}
\newcounter{minutes}

\newcommand{\met}{\mbox{${E\!\!\!\!/_{\rm T}}$}}

\newcommand{\et}{\mbox{$E_{{\rm T}}$}}

\def \sp       {\relax\ifmmode{\;}\else{$\;$}\fi}	
\def\Journal#1#2#3#4{{#1} {\bf #2}, #3 (#4)}
\def \PRL      {Phys. Rev. Lett.~}
\def \PR       {Phys. Rev.}
\def \PRD      {Phys. Rev. D}

\def \PLB      {Phys. Lett. B}
\def \ZPC      {Z. Phys. C}	
\def \NPB      {Nucl. Phys. B}
\def \PR       {Phys. Rep.~}

\def \NIM      {Nucl. Instrum. Methods}
\def \NIMA     {Nucl. Instrum. Methods Phys. Res. Sect. A}
\def \RPP	{Rep. Prog. Phys.}
\def \PRP	{Prog. Theor. Phys}
\def \MPL	{{Mod. Phys. Lett.} A}
\def \EPJC	{{Eur. Phys. J.} C}
\def \CPC	{Comput. Phys. Commun.}
\begin{document}
\title
{A New Mass Reconstruction Technique for Resonances Decaying to $\tau\tau$}

\author[tamu]{A.Elagin}
\author[fnal]{P.Murat}
\author[lbnl]{A.Pranko}
\author[tamu]{A.Safonov}
\address[tamu]{Department of Physics and Astronomy, Texas A\&M Universitiy, College Station, TX 77843, USA}
\address[fnal]{Fermi National Accelerator Laboratory, Batavia, IL 60506, USA}
\address[lbnl]{Ernest Orlando Lawrence Berkeley National Laboratory, Berkeley, CA 94720, USA}

\date{\today}

\begin{abstract}
Accurate reconstruction of the mass of a resonance decaying to a pair
of $\tau$ leptons is challenging because of the presence of multiple 
neutrinos from $\tau$ decays. The existing methods rely on either a 
partially reconstructed mass, which has a broad spectrum that reduces 
sensitivity, or the collinear approximation, which is applicable only 
to the relatively small fraction of events. We describe a new technique, 
which provides an accurate mass reconstruction of the original resonance 
and does not suffer from the limitations of the collinear approximation. 
The major improvement comes from replacing assumptions of the collinear 
approximation by a requirement that mutual orientations of the neutrinos 
and other decay products are consistent with the mass and decay kinematics 
of a $\tau$ lepton. This is achieved by minimizing a likelihood function 
defined in the kinematically allowed phase space region. In this paper 
we describe the technique and illustrate its performance using 
$Z/\gamma^{*}\to\tau\tau$ and $H\to\tau\tau$ events simulated with the 
realistic detector resolution. The method is also tested on a clean
sample of data $Z/\gamma^{*}\to\tau\tau$ events collected by the CDF 
experiment at the Tevatron. We expect that this new technique will
allow for a major improvement in searches for the Higgs boson at
both the LHC and the Tevatron.

\end{abstract}


\maketitle

\section{Introduction}
\label{sec:introduction}

Invariant mass reconstruction is commonly used in experimental searches for new physics, such as for the 
Higgs or Z$^\prime$ bosons, as well as in measurements of properties of known resonances. This technique is 
relatively straightforward for $e^+e^-$, $\mu^+\mu^-$, or di-jet final states. The accuracy of mass 
reconstruction in these channels is dominated by the detector resolution for lepton or jet momenta.
The sensitivity of ``mass bump-hunting'' analyses depends critically on how narrow the signal invariant 
mass distribution is compared to the (usually broad) distributions in background processes. Unfortunately,
this simple strategy is much less effective in searches for resonances decaying to a pair of $\tau$ leptons 
because the $\tau$ lepton energy associated with neutrinos escapes detection, and only visible products 
(leptons in the case of leptonic $\tau$ decays or jets in the case of hadronic $\tau$ decays) are observed 
in the detector.

\begin{figure*}[htb]
\includegraphics[width=0.5\linewidth]{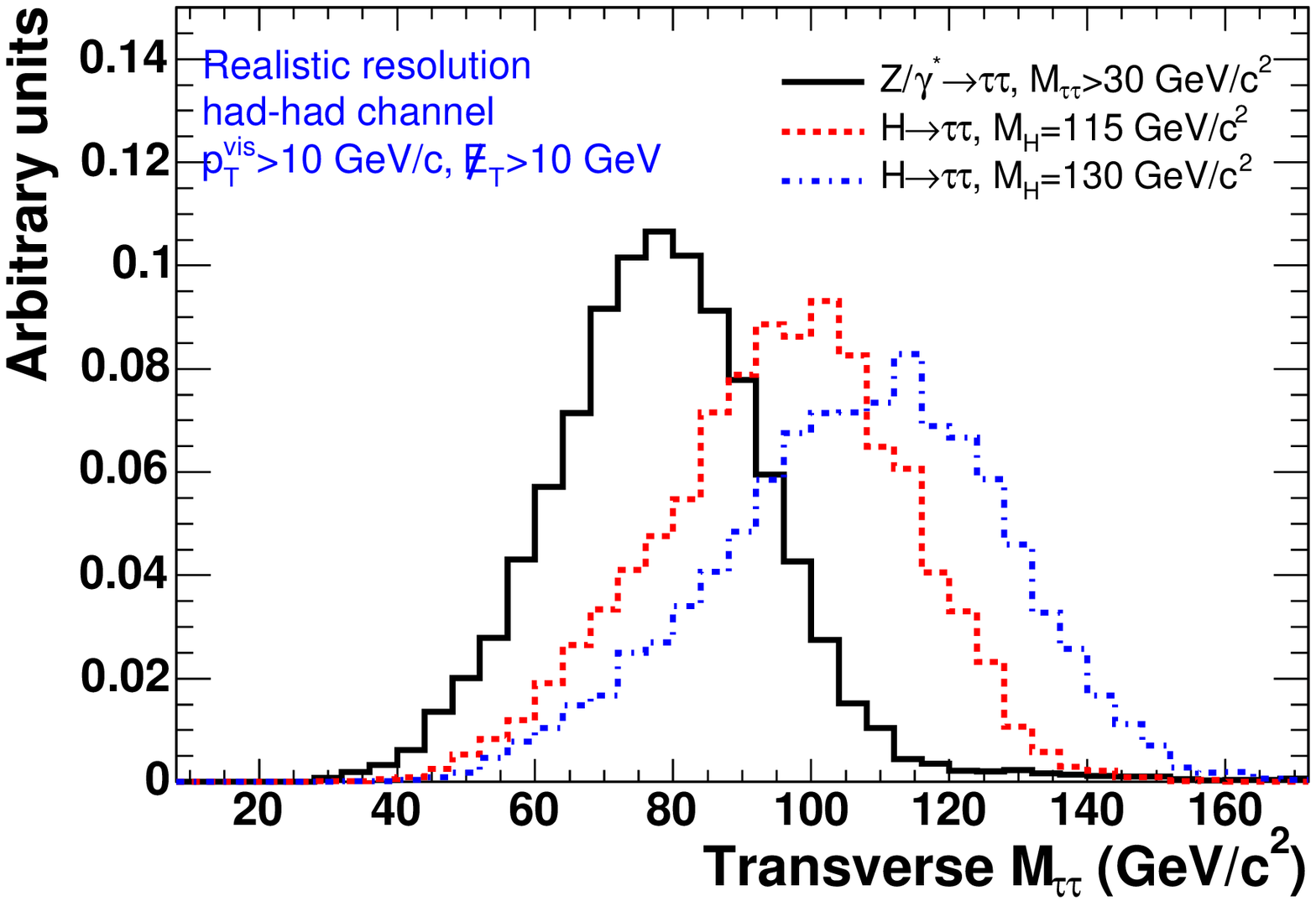}
\includegraphics[width=0.5\linewidth]{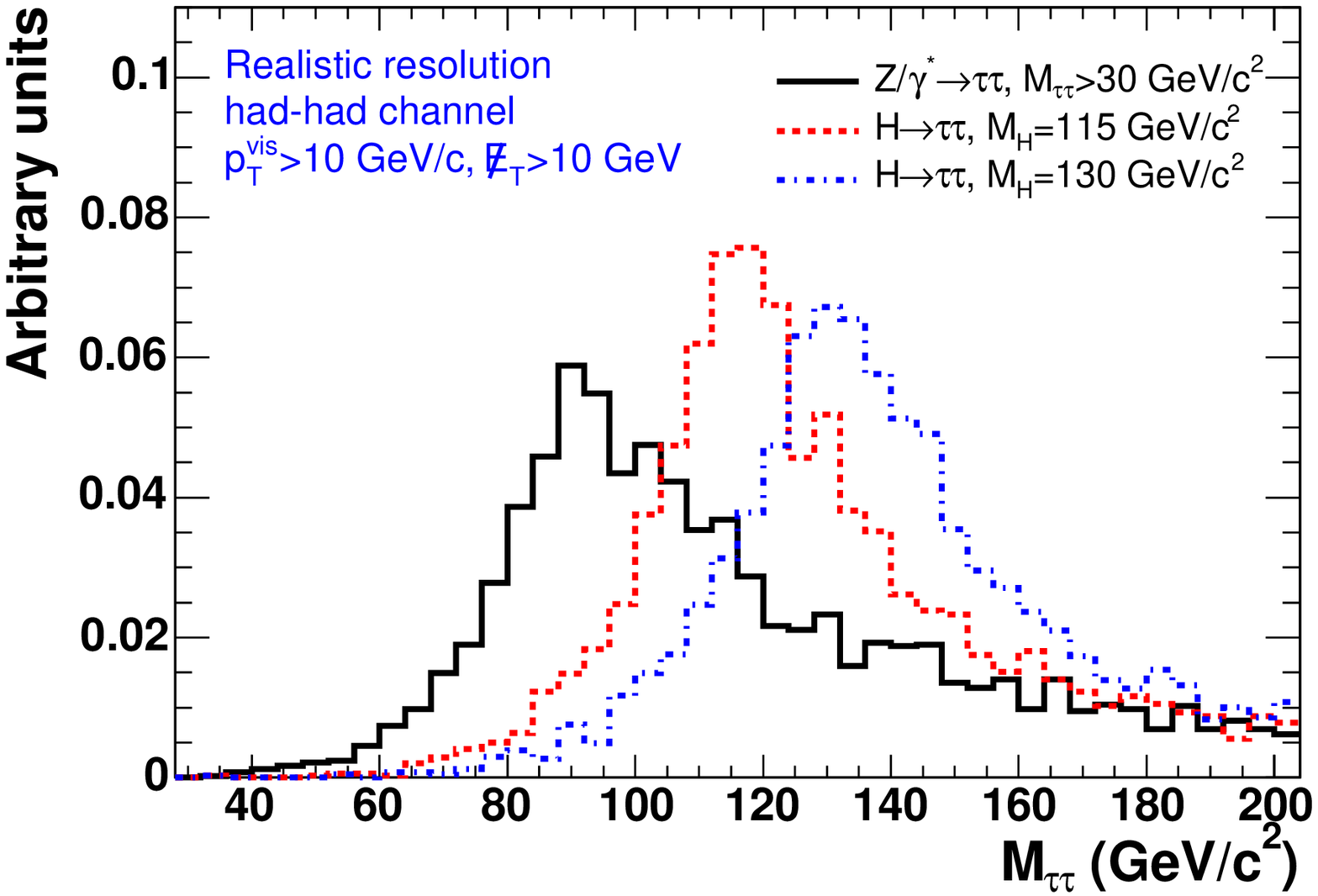}
\caption{Example of the transverse mass (left plot) defined as an invariant mass of $\met$ and 
visible $\tau$ decay products, and the fully reconstructed mass (right plot) using the collinear 
approximation for three event samples: inclusive $Z/\gamma^{*}\to\tau\tau$ and $gg\to H\to\tau\tau$ with 
$M_H$=115 and 130~GeV/c$^2$. Results are obtained for the fully hadronic $\tau\tau$ decay mode. Events 
are simulated with the realistic detector resolution (discussed in Sec.~\ref{sec:MMCdetector}). All 
distributions are normalized to the unit area.  
\label{fig:cdf_mass}}
\end{figure*}

Each $\tau$ lepton decay involves one or two neutrinos, depending on the final state: hadronic ($\tau\to\nu_{\tau}+hadrons$) or leptonic ($\tau\to\nu_{\tau}+l\bar{\nu_l}$, where $l$=$e$ or $\mu$). In $pp$ or $p\bar{p}$ collisions, the full energy of neutrinos cannot be determined. Instead, one can only reconstruct a 
transverse energy imbalance in the calorimeter (or missing transverse energy, $\met$), which is representative 
of the total transverse momentum of all neutrinos in the event. Therefore, when two or more neutrinos are produced 
in the same event, their individual transverse momenta and directions cannot be reconstructed. The situation in 
decays of heavy resonances into two $\tau$ leptons is even more complex. In these events, the two $\tau$'s are 
often produced ``back-to-back'' and the missing momentum associated with their neutrinos partially cancels out. 
As a result, the invariant mass of a resonance cannot be directly reconstructed from the $\met$ and visible 
decay products of $\tau$ leptons. Various techniques exist to partially reconstruct the mass of resonances in 
$\tau\tau$ final states. However, the reconstructed mass distributions for signal processes are rather broad 
(with long tails and typical core resolutions on the order of $\sim$20$\%$), which makes it difficult to separate 
them from the background and considerably reduces the signal significance. This poses a major challenge for the Higgs boson searches in the $H\to\tau\tau$ channel, one of the most important channels for discovering a low-mass Higgs 
boson at the LHC~\cite{atlas_tdr,cms_tdr}, whether in the context of the Standard Model or beyond (for example, in supersymmetric models). Another challenge in searching for a low-mass Higgs boson in the $\tau\tau$ channel is the large and irreducible background from $Z/\gamma^{*}\to\tau\tau$ events. This is because the $Z/\gamma^{*}$ background is several orders of magnitude larger than any expected Higgs signal, and its broad partially reconstructed mass distribution completely dominates the signal region (for example, see Fig.~\ref{fig:cdf_mass} or reference~\cite{cdf_higgs_prl}). Therefore, a major improvement in $\tau\tau$ invariant mass, $M_{\tau\tau}$, reconstruction techniques is needed in order to significantly enhance the sensitivity of $H\to\tau\tau$ searches 
at the Tevatron and LHC experiments.    

In this paper, we propose a new method, which substantially improves the accuracy of the $\tau\tau$ invariant mass reconstruction. We expect it will lead to a major improvement in the sensitivity of the Higgs boson searches in the $H\to\tau\tau$ channel at the Tevatron and LHC. In the next section, we briefly review currently used methods. Section~\ref{sec:MMC} describes the new technique and illustrates its performance using a Monte Carlo simulation with the realistic detector resolution. In Sec.~\ref{sec:Data}, we report the results of tests on a clean sample of data $Z/\gamma^{*}\to\tau\tau$ events collected by the CDF experiment at the Tevatron. Finally, we conclude in Sec.~\ref{sec:Conclusions}.

\section{Review of the Commonly Used Techniques for $\tau\tau$ Mass Reconstruction}
\label{sec:review}

The two methods frequently used at hadron colliders either rely on reconstructing a partial invariant mass 
or use the collinear approximation. In this section, we review these techniques and discuss their 
advantages and shortcomings.

\subsection{The Transverse Mass Method}
\label{sec:vismass}

Neutrinos from the $\tau$ decays escape detection and make it impossible to determine the 4-momenta of 
$\tau$ leptons and thus $M_{\tau\tau}$. Therefore, one of the simplest and frequently used methods relies 
on a partial (or reduced) invariant mass reconstruction. Examples include either the invariant mass of 
visible decay products of the two $\tau$ leptons or the invariant mass of the visible decay products and 
$\met$ in the event, transverse mass. The latter is defined as follows: 
\begin{eqnarray}
M^2(\tau_{\mathrm{vis}_1},\tau_{\mathrm{vis}_2},\met)=m^2_{\mathrm{vis}_1} + m^2_{\mathrm{vis}_2} + \nonumber \\
2\times\left(\sqrt{m^2_{\mathrm{vis}_1}+p^2_{\mathrm{vis}_1}}\sqrt{m^2_{\mathrm{vis}_2}+p^2_{\mathrm{vis}_2}}
+\met\sqrt{m^2_{\mathrm{vis}_1}+p^2_{\mathrm{vis}_1}}+\met\sqrt{m^2_{\mathrm{vis}_2}+p^2_{\mathrm{vis}_2}}\right) - \nonumber \\
2\times(\vec{p}_{\mathrm{vis}_1}\cdot\vec{p}_{\mathrm{vis}_2}+\vec{p}_{\mathrm{vis}_1}\cdot\vec{\met}
+\vec{p}_{\mathrm{vis}_2}\cdot\vec{\met}), 
\label{eq:math_mvisT}
\end{eqnarray}
where $\vec{\met}$=$(\met_x,\met_y,0)$ and $\vec{p}_{\mathrm{vis}_{1,2}}$, $m_{\mathrm{vis}_{1,2}}$ are the 
momenta vectors and invariant masses of the visible $\tau$ decay products. The transverse mass provides 
a somewhat better separation from the QCD multi-jet backgrounds with fake $\tau$ signatures, and it is 
often preferred in data analyses. The advantage of this technique is that the partial mass can be defined 
for all signal events, thus preserving the statistical power of the available data. However, ignoring 
or not fully accounting for the neutrino momenta biases and broadens the reconstructed $M_{\tau\tau}$
distributions, and therefore leads to a significantly reduced sensitivity in searches and measurements.
This problem is particularly prominent in the low-mass $H\to\tau\tau$ search, where the signal cannot be 
separated from the much larger and very broad $Z\to\tau\tau$ background. This is illustrated in the left
plot in Fig.~\ref{fig:cdf_mass}, which shows the transverse mass $M(\tau_{\mathrm{vis}_1}, \tau_{\mathrm{vis}_2},\met)$ distribution.

\subsection{Collinear Approximation Technique}
\label{sec:CollApp}

The collinear approximation is another frequently used technique~\cite{atlas_tdr,cms_tdr}. This method was 
first proposed in reference~\cite{collinear_approx} to reconstruct the invariant mass in $\tau\tau$ decays 
of a Higgs boson produced in association with an energetic jet. It is based on two important assumptions: 
that the neutrinos from each $\tau$ decay are nearly collinear with the corresponding visible $\tau$ decay 
products (i.e., $\phi_\nu$$\simeq$$\phi_{\mathrm{vis}}$ and $\theta_\nu$$\simeq$$\theta_{\mathrm{vis}}$); and that the 
$\met$ in the event is due only to neutrinos. In this case, the total invisible momentum carried away by 
neutrinos in each $\tau$ decay can be estimated by solving two equations:
\begin{eqnarray}
\met_x = p_{\mathrm{mis}_1} \sin{\theta_{\mathrm{vis}_1}} \cos{\phi_{\mathrm{vis}_1}} +  p_{\mathrm{mis}_2} \sin{\theta_{\mathrm{vis}_2}} \cos{\phi_{\mathrm{vis}_2}}	\nonumber\\
\met_y = p_{\mathrm{mis}_1} \sin{\theta_{\mathrm{vis}_1}} \sin{\phi_{\mathrm{vis}_1}} +  p_{\mathrm{mis}_2} \sin{\theta_{\mathrm{vis}_2}} \sin{\phi_{\mathrm{vis}_2}},	
\label{eq:math_collinear_approximation}
\end{eqnarray} 
where $\met_x$ and $\met_y$ are the $x$- and $y$-components of the $\met$ vector, $p_{\mathrm{mis}_1}$ and $p_{\mathrm{mis}_2}$ are the combined invisible momenta (there can be two $\nu$'s in a $\tau$ decay) of each $\tau$ decay, and $\theta_{\mathrm{vis}_{1,2}}$ and $\phi_{\mathrm{vis}_{1,2}}$ are the polar and azimuthal angles of
the visible products of each $\tau$ decay. Then, the invariant mass of the $\tau\tau$-system can be calculated 
as $M_{\tau \tau}$=$m_{\mathrm{vis}}/\sqrt{x_1 x_2}$ , where $m_{\mathrm{vis}}$ is the invariant mass of visible $\tau$ decay products, and $x_{1,2}$=$p_{\mathrm{vis}_{1,2}}/(p_{\mathrm{vis}_{1,2}}+p_{\mathrm{mis}_{1,2}})$ are momentum fractions carried away by visible $\tau$ decay products. Despite offering the great advantage of a fully reconstructed $\tau\tau$ mass ($M_{\tau\tau}$) instead of a partial visible mass, the collinear approximation still
has significant shortcomings. The technique gives a reasonable mass resolution only for the small fraction of 
events where the $\tau\tau$ system is boosted, i.e., produced in association with a large $\et$ jet, and the visible $\tau$ decay products are not back-to-back in the plane transverse to the beam line. The last requirement is needed, because the system of Eqs.~\ref{eq:math_collinear_approximation} becomes degenerate if $\phi_{\mathrm{vis}_1}$=$\phi_{\mathrm{vis}_2}+\pi$ and solutions $p_{\mathrm{mis}_{1,2}}$$\sim$$\sin^{-1}(\phi_{\mathrm{vis}_1}-\phi_{\mathrm{vis}_2})$ diverge as $|\phi_{\mathrm{vis}_1}-\phi_{\mathrm{vis}_2}|\to\pi$. 
Unfortunately, the majority of $H\to\tau\tau$ events are produced with $\tau$ leptons in nearly the back-to-back topology. Therefore, this technique is applicable only to a relatively small fraction of $\tau\tau$ events. The collinear approximation is also very sensitive to the $\met$ resolution and tends to over-estimate the $\tau\tau$ mass, leading to long tails in the reconstructed mass distribution (see right plot in Fig.~\ref{fig:cdf_mass}). This effect is especially undesirable for low-mass Higgs boson searches, where the tails of a much larger $Z\to\tau\tau$ background completely overwhelm the expected Higgs peak region.  

\section{The Missing Mass Calculator Technique}
\label{sec:MMC}

The new technique proposed in this paper, the Missing Mass Calculator (MMC) method, allows for a complete reconstruction of event kinematics in the $\tau\tau$ final states with significantly improved invariant mass and neutrino momentum resolutions. The MMC technique does not suffer from the limitations of the collinear approximation described in the previous section and can be applied to all $\tau\tau$ event topologies without sacrificing the reconstructed mass resolution.

\subsection{The Concept and Method Description}
\label{sec:MMCconcept}

To facilitate the description of the method, we begin with assuming a perfect detector resolution and that there are
no other neutrinos in $\tau\tau$ events except for those from the $\tau$ lepton decays. Under these assumptions, full reconstruction of the event topology requires solving for 6 to 8 unknowns: $x$-, $y$-, and $z$-components of the 
invisible momentum carried away by neutrino(s) for each of the two $\tau$ leptons in the event, and, if one or both 
$\tau$'s decay leptonically, the invariant mass of the neutrinos from each leptonic $\tau$ decay. However, there are
only 4 equations connecting these unknowns:
\begin{eqnarray}
\met_x = p_{\mathrm{mis}_1} \sin{\theta_{\mathrm{mis}_1}} \cos{\phi_{\mathrm{mis}_1}}+  p_{\mathrm{mis}_2} \sin{\theta_{\mathrm{mis}_2}} \cos{\phi_{\mathrm{mis}_2}}	\nonumber \\
\met_y = p_{\mathrm{mis}_1} \sin{\theta_{\mathrm{mis}_1}} \sin{\phi_{\mathrm{mis}_1}} +  p_{\mathrm{mis}_2} \sin{\theta_{\mathrm{mis}_2}} \sin{\phi_{\mathrm{mis}_2}}	\nonumber \\
M_{\tau_1}^2 = m^2_{\mathrm{mis}_1} + m^2_{\mathrm{vis}_1}+ 2 \sqrt{p_{\mathrm{vis}_1}^2+ m^2_{\mathrm{vis}_1}}\sqrt{p_{\mathrm{mis}_1}^2+ m^2_{\mathrm{mis}_1}}\nonumber \\ 
-2 p_{\mathrm{vis}_1} p_{\mathrm{mis}_1} \cos{\Delta \theta_{vm_1}}  \nonumber \\
M_{\tau_2}^2 = m^2_{\mathrm{mis}_2} + m^2_{\mathrm{vis}_2}+ 2 \sqrt{p_{\mathrm{vis}_2}^2+ m^2_{\mathrm{vis}_2}}\sqrt{p_{\mathrm{mis}_2}^2+ m^2_{\mathrm{mis}_2}} \nonumber \\ 
-2 p_{\mathrm{vis}_2} p_{\mathrm{mis}_2} \cos{\Delta \theta_{vm_2}}
\label{eq:math_mms}
\end{eqnarray}   
where $\met_x$ and $\met_y$ are the $x$- and $y$-components of the $\met$ vector, $p_{\mathrm{vis}_{1,2}}$, $m_{\mathrm{vis}_{1,2}}$, $\theta_{\mathrm{vis}_{1,2}}$, $\phi_{\mathrm{vis}_{1,2}}$ are the momenta, invariant masses, polar and azimuthal angles of the visible $\tau$ decay products, and $M_{\tau}$=1.777~GeV/c$^2$ is the $\tau$ lepton invariant mass. 
The rest of the variables constitute the ``unknowns'' which are the combined invisible (``missing'') momenta $\vec{p}_{\mathrm{mis}_{1,2}}$ carried away by the neutrino (or neutrinos) for each of the two decaying $\tau$ leptons and the invariant mass of the neutrino(s) in the $\tau$ decay, $m_{\mathrm{mis}_{1,2}}$. Finally, $\delta \theta_{vm_{1,2}}$ is the angle between the vectors $p_{\mathrm{mis}}$ and $p_{\mathrm{vis}}$ for each of the two $\tau$ leptons, and it can be expressed in terms of the other variables. For hadronic decays of $\tau$'s, the $m_{\mathrm{mis}}$ is set to 0 as there is only one neutrino 
involved in the decay. This reduces the number of unknowns.

\begin{figure*}[htb]
\includegraphics[width=0.31\linewidth]{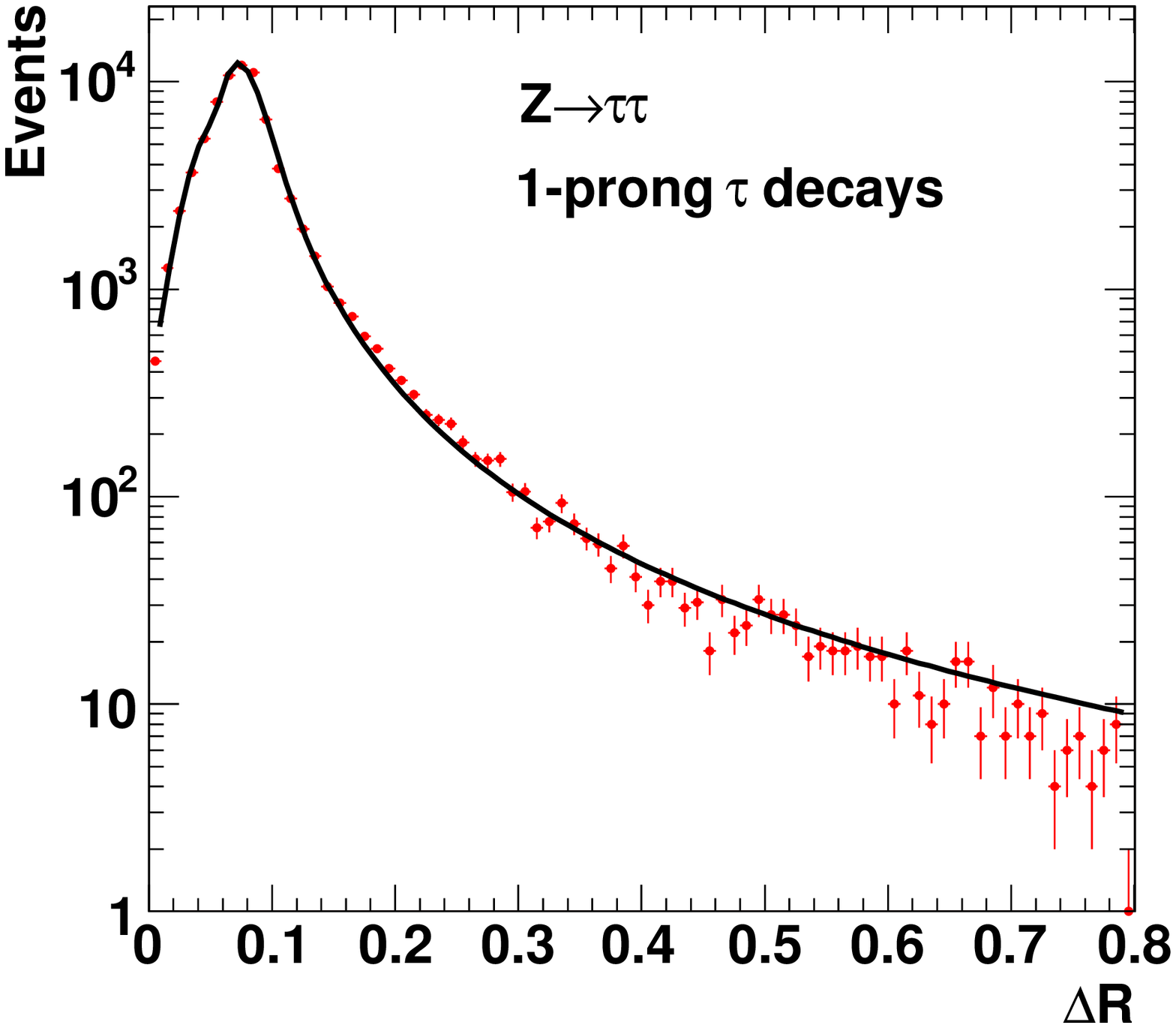}
\includegraphics[width=0.31\linewidth]{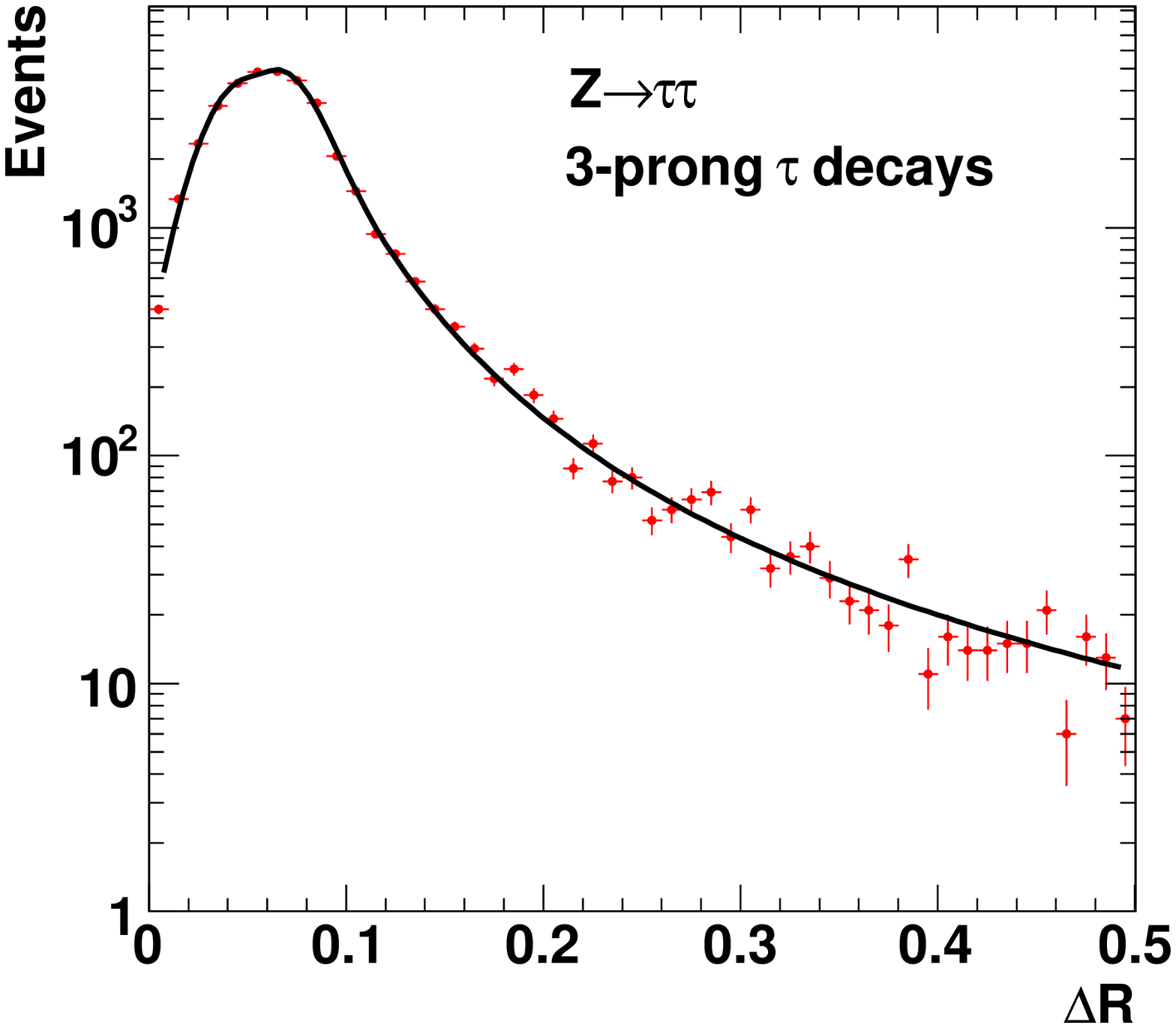}
\includegraphics[width=0.31\linewidth]{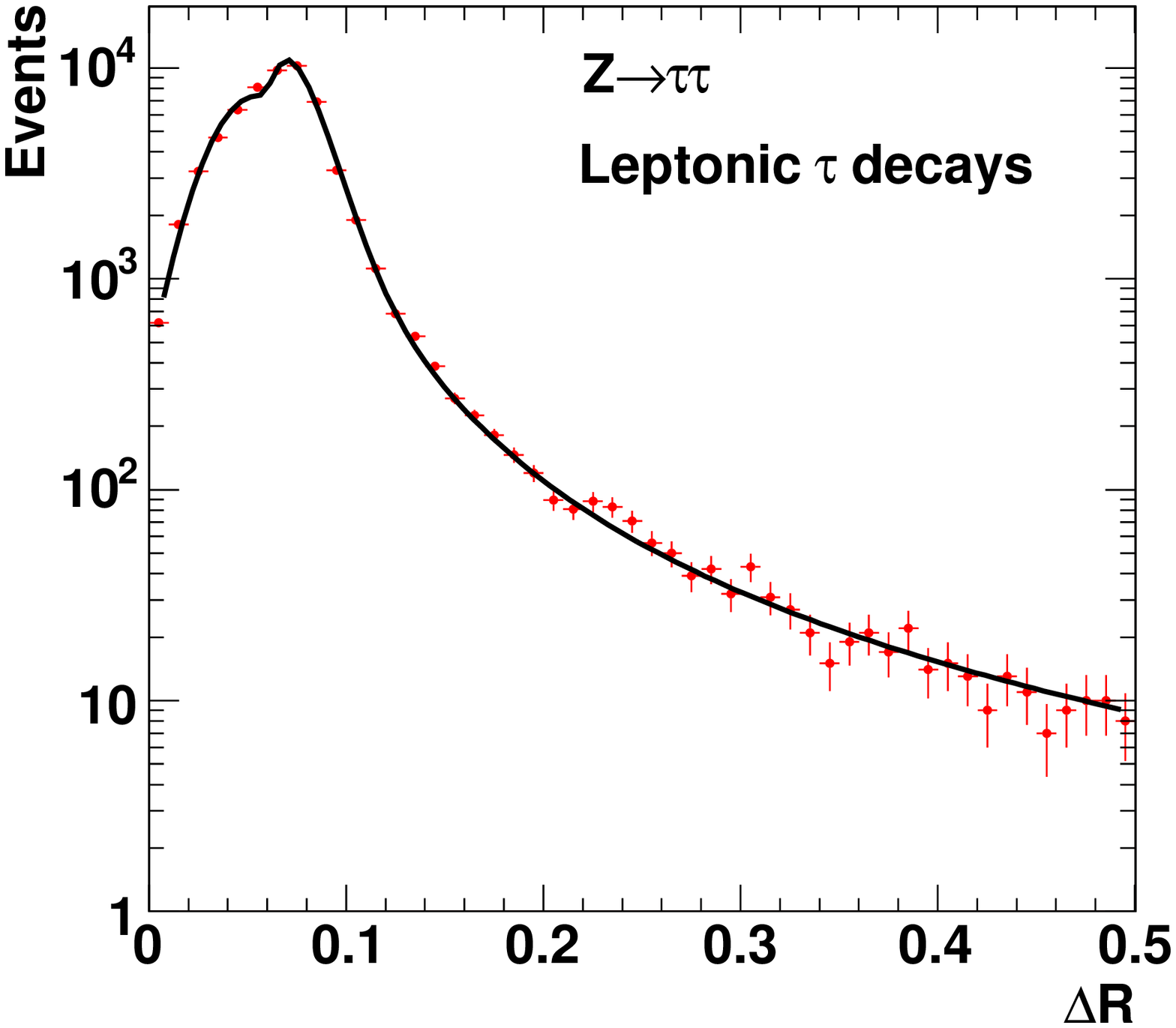}
\caption{Example of the probability distribution functions ${\cal P}(\Delta R,p_{\tau})$ for a particular value 
of the original $\tau$ lepton momentum ($p_{\tau}$). These functions are used in the calculation of the likelihood 
$\cal L$ for three cases: 1-prong $\tau$ (left plot), 3-prong $\tau$ (middle plot), and leptonic decays (right plot) 
of $\tau$ leptons.  These distributions depend only on the decay type and initial momentum of the $\tau$ lepton.
}
\label{fig:mmc_mmc_dR}
\end{figure*}

The number of unknowns (from 6 to 8, depending on the number of leptonic $\tau$ decays) exceeds the number of constraints. Therefore, the available information is not sufficient to find the exact solution. However, not all solutions of this under-constrained system are equally likely, and additional knowledge of $\tau$ decay kinematics can be used to distinguish more likely solutions from less likely ones. An example of such additional information is the expected angular distance between the neutrino(s) and the visible decays products of the $\tau$ lepton. Figure~\ref{fig:mmc_mmc_dR} shows the distribution for the distance 
$\Delta R$=$\sqrt{(\eta_{\mathrm{vis}}-\eta_{\mathrm{mis}})^2 + (\phi_{\mathrm{vis}}-\phi_{\mathrm{mis}})^2}$ between the directions of visible and invisible (missing) decay products\footnote{For simplicity, we use the $\Delta R$ parametrization,
although a 3-dimensional angle between the decay products might be a more natural choice.} for the three distinct $\tau$ decay types: leptonic, 1-prong hadronic and 3-prong hadronic. We incorporate this additional knowledge of decay kinematics as probability density functions in a properly defined global event likelihood to provide additional constraints and obtain a better estimator of $M_{\tau\tau}$.

\begin{figure*}[htb]
\includegraphics[width=0.50\linewidth]{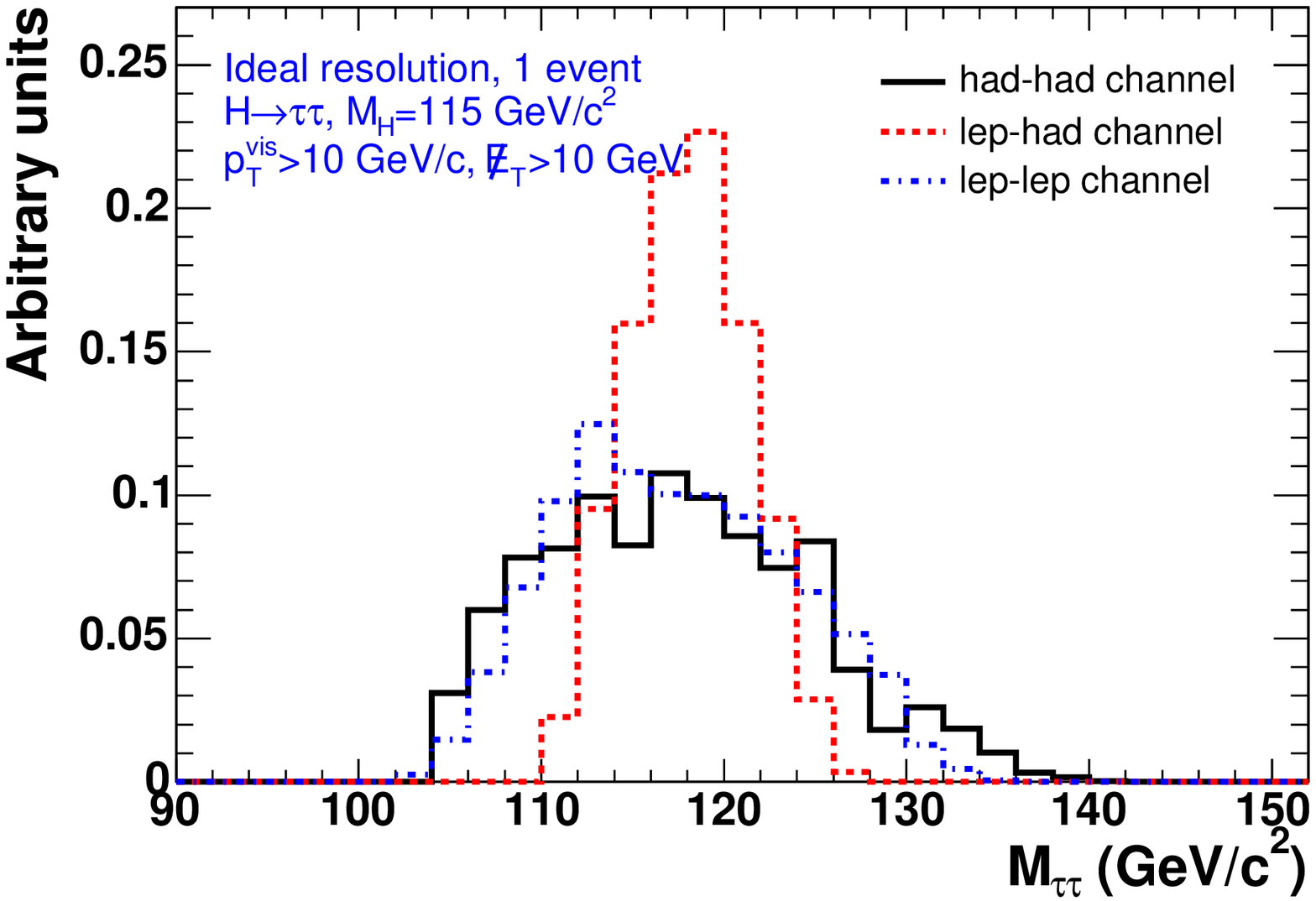}
\includegraphics[width=0.50\linewidth]{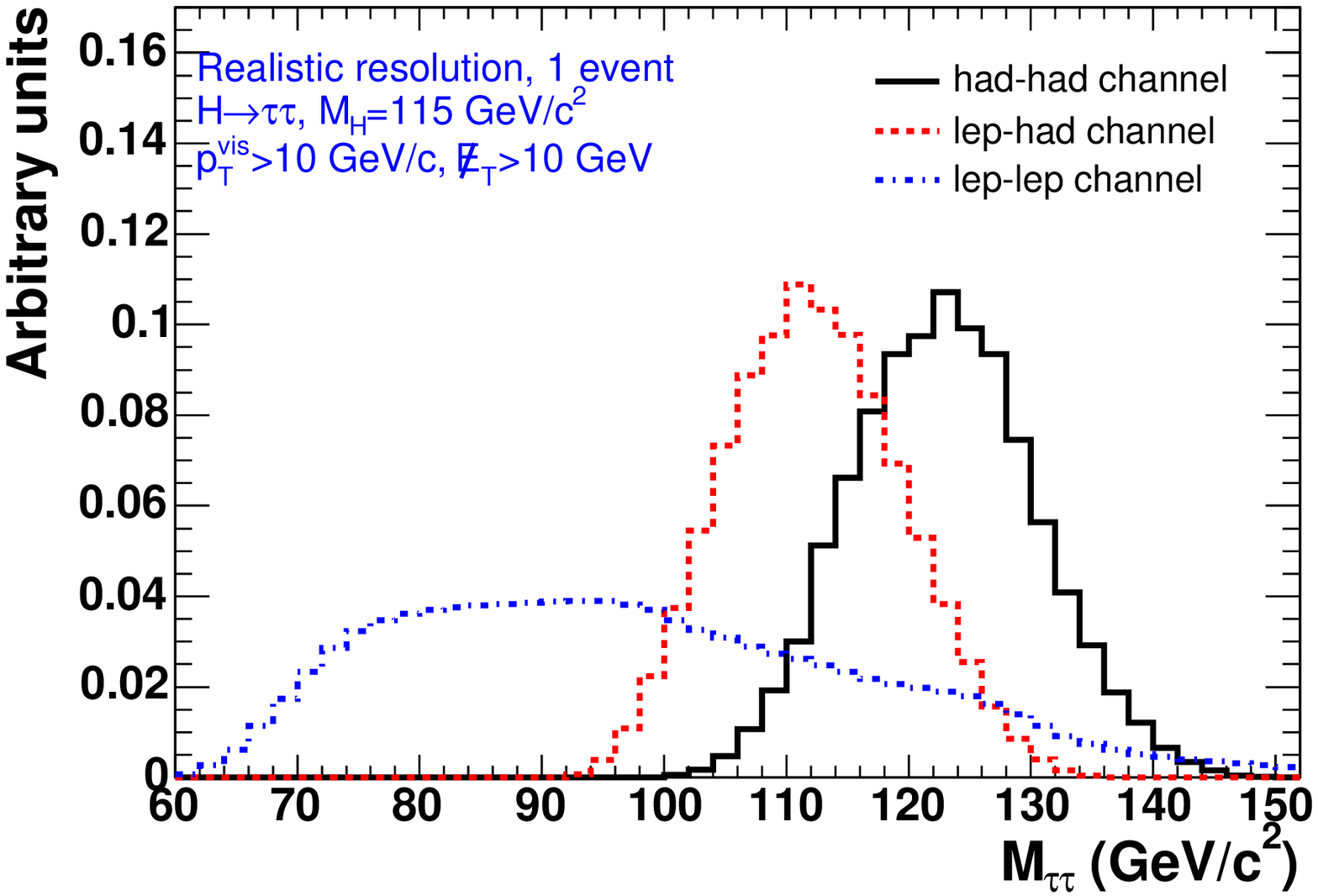}
\caption{Example of the $M_{\tau \tau}$ distribution filled for all grid points in one of the $H\to\tau\tau$ events for each of the three decay modes. An entry for each point is weighted by its $\cal L$. Plot on the left shows $M_{\tau \tau}$ for the case of the ideal detector resolution and plot on the right demonstrates $M_{\tau \tau}$ for the same three events in the case of the realistic
detector resolution. All distributions are normalized to a unit area.}
\label{fig:mmc_example_1event}
\end{figure*}

We first describe the method for the better constrained case, where both $\tau$'s decay hadronically, and then we explain how 
the machinery is adjusted for the case of leptonic decays. When both $\tau$'s decay hadronically, the system of Eqs.~\ref{eq:math_mms} can be solved exactly for any point in, for example, the ($\phi_{\mathrm{mis}_1}$, $\phi_{\mathrm{mis}_2}$) parameter space. For each point in that grid, the vectors $p_{\mathrm{mis}_{1,2}}$ are fully defined and, therefore, one can calculate the distance $\Delta R_{1,2}$ between the vector $p_{\mathrm{vis}_{1,2}}$ and the current assumed direction of $p_{\mathrm{mis}_{1,2}}$. To evaluate the probability of such decay topology, we use $\Delta R$ distributions similar to those shown in Fig.~\ref{fig:mmc_mmc_dR}, but we take into account the dependence of the distribution on the momentum of the initial $\tau$ lepton. If the $\tau$ lepton polarization is neglected, the $\Delta R$ distribution depends only on the $\tau$ momentum 
and decay type, but not on the source of $\tau$'s. Therefore, we use simulated $Z/\gamma^{*}\to\tau\tau$ events to obtain $\Delta R$ distributions for small bins (5 GeV/$c$) in the initial $\tau$ momentum, $p$, in the range 10~GeV/$c$$<$$p$$<$100~GeV/$c$ (the range can be extended to both smaller and larger values). Events are simulated using Pythia~\cite{pythia} supplemented with the TAUOLA package~\cite{tauola} for $\tau$ decays. To simplify the calculations further, we parametrize the $\Delta R$ distributions by fitting them with a linear combination of Gaussian and Landau functions. Examples of such fits are shown as solid lines in Fig.~\ref{fig:mmc_mmc_dR}. The $p_{\tau}$-dependence of the mean, width and relative normalization of the Gaussian and Landau is then parametrized as $p_0/(x+p_1x^2)+p_2+p_3x+p_4x^2$, yielding fully parametrized distributions ${\cal P}(\Delta R, p)$, which can be used to evaluate the probability of a particular $\tau$ decay topology. To incorporate this information as an additional constraint, we define the logarithm of the event probability (or likelihood) as follows:
\begin{eqnarray}
{\cal L}=-\log{({{\cal P}(\Delta R_1,p_{\tau1})}\times{{\cal P}(\Delta R_2,p_{\tau1})})},
\label{eq:likelihood}
\end{eqnarray}
where functions $\cal P$ are chosen according to one of the decay types. To determine the best estimate for the $\tau\tau$ invariant mass in a given event, we produce an $M_{\tau \tau}$ distribution for all scanned points in the $(\phi_{\mathrm{mis}_1}$, $\phi_{\mathrm{mis}_2})$ grid weighed by a corresponding probability, ${{\cal P}(\Delta R_1,p_{\tau1})}\times{{\cal P}(\Delta R_2,p_{\tau1})}$. The most probable value of the $M_{\tau \tau}$ distribution is used as the final estimator of $M_{\tau\tau}$ for a given event. An example of a such $M_{\tau \tau}$ histogram for typical $H\to\tau\tau$ events of each category is shown in Fig.~\ref{fig:mmc_example_1event}. A similar procedure can also be used to build estimators for other kinematic variables, if desired.

For events where one or both $\tau$ leptons decay leptonically, the above procedure is adjusted to account for the unknown value of $m_{\mathrm{mis}}$ of the two neutrinos in each of the leptonically decaying $\tau$'s in the event. In this case,
the scan is performed in a phase space of higher dimensionality: ($\phi_{\mathrm{mis}_1}$,$\phi_{\mathrm{mis}_2}$,$m_{\mathrm{mis}_1}$), if only one of the two $\tau$'s decay leptonically; or ($\phi_{\mathrm{mis}_1}$,$\phi_{\mathrm{mis}_2}$,$m_{\mathrm{mis}_1}$,$m_{\mathrm{mis}_2}$), if both decay $\tau$'s decay to leptons. As in the fully hadronic mode, one can unambiguously reconstruct the 4-momenta of both $\tau$ leptons for each point on the grid and calculate the event probability according to Eq.~\ref{eq:likelihood}. For simplicity, we scan uniformly in the entire range of kinematically allowed values of $m_{\mathrm{mis}}$, but a scan performed according to the $m_{\mathrm{mis}}$ probability distribution function obtained from simulation may improve the algorithm performance.

\subsection{Performance of the MMC Technique with Ideal Detector Resolution}
\label{sec:MMCideal}

\begin{figure*}[htb]
\includegraphics[width=0.50\linewidth]{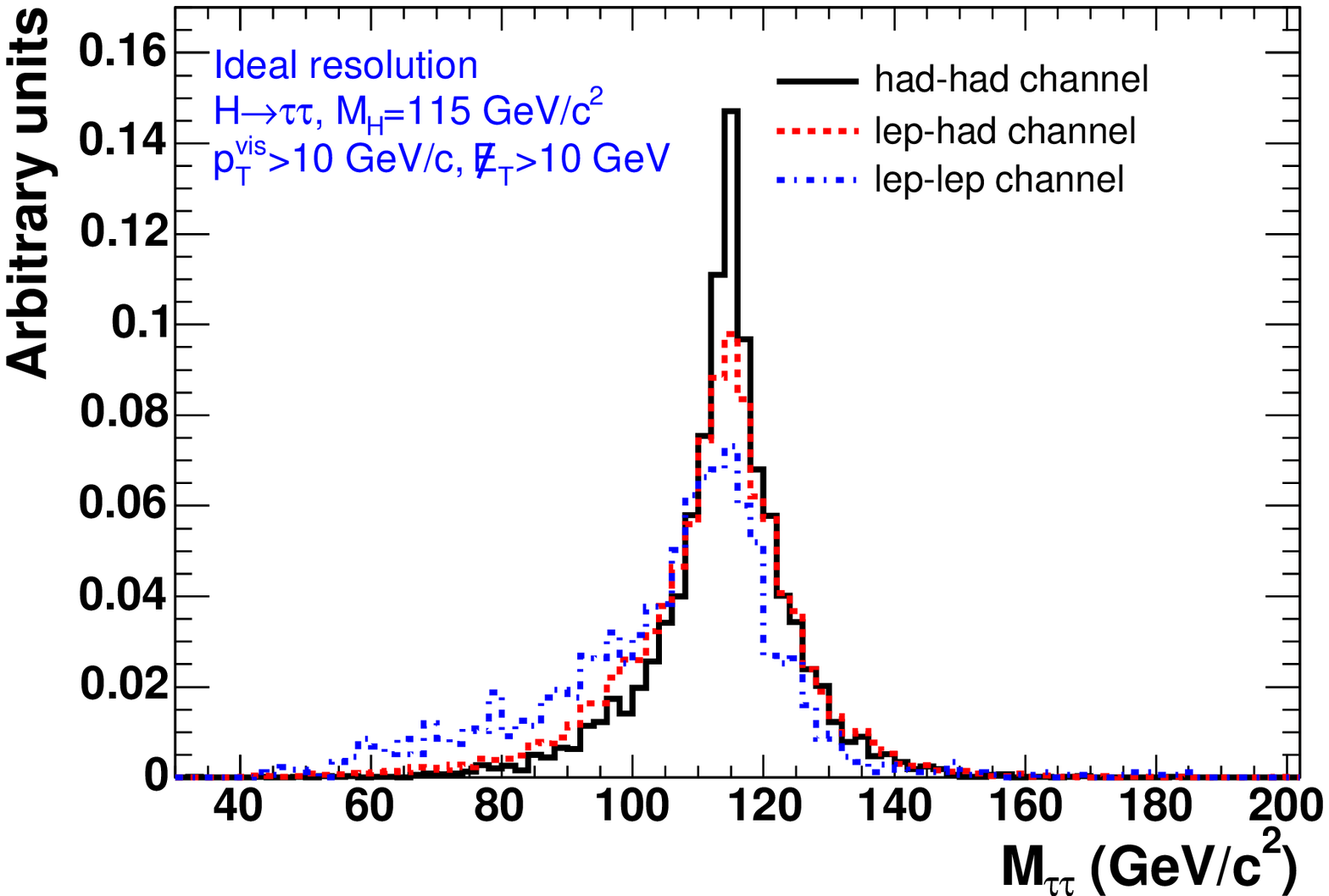}
\includegraphics[width=0.50\linewidth]{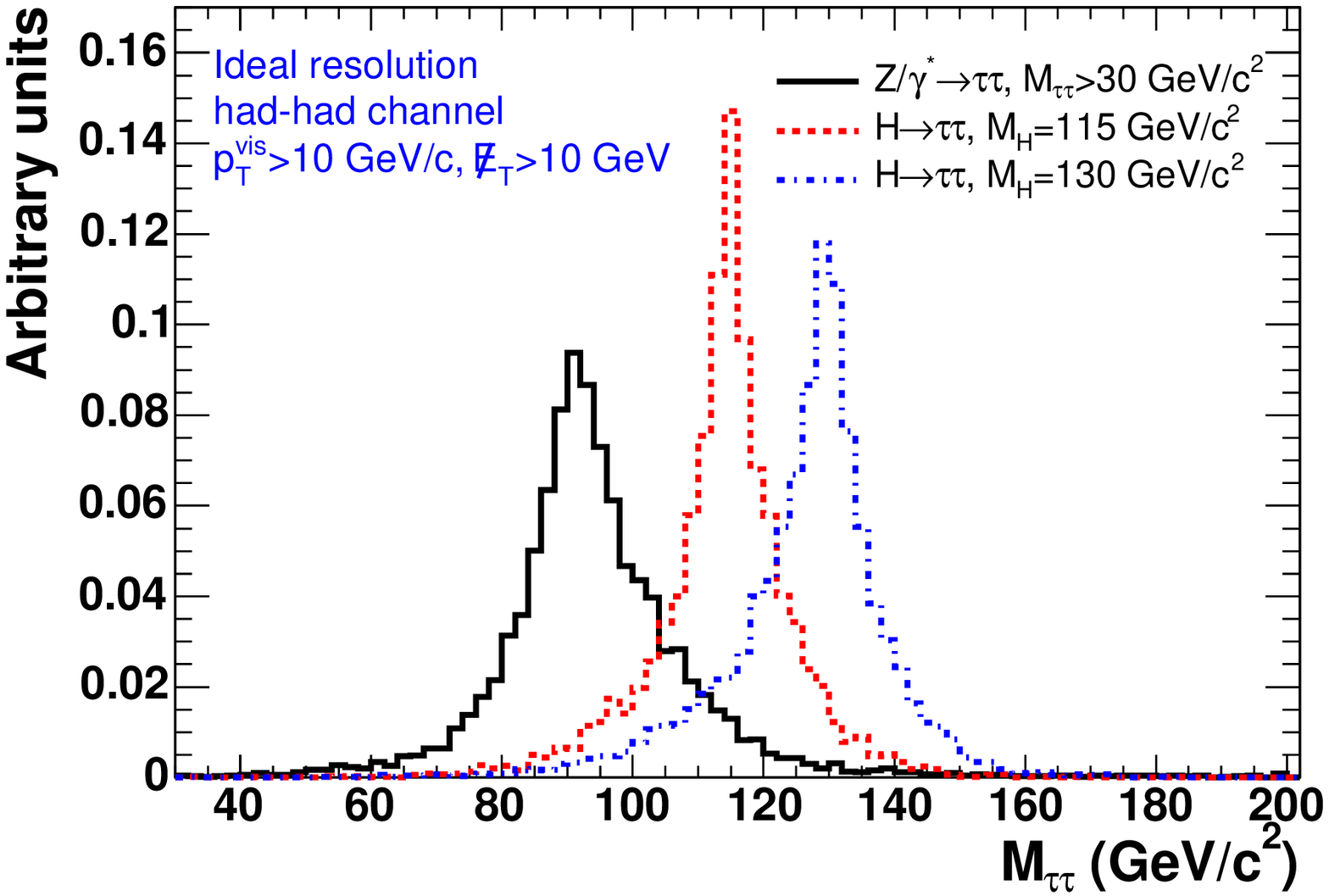}
\caption{Left plot demonstrates the reconstructed $M_{\tau\tau}$ in $H\to\tau\tau$ events with $M_H=$115~GeV/c$^2$
for each of the three decay categories: both $\tau$'s decay hadronically (solid line); one $\tau$ decays leptonically 
and the other one hadronically (dashed line); and both $\tau$'s decay leptonically (dashed-dotted line). Right plot shows 
the reconstructed mass in $Z/\gamma^{*}\to\tau\tau$ and $H\to\tau\tau$ events with $M_H=$115 and 130~GeV/c$^{2}$ in the fully 
hadronic decay mode. Results are obtained in the assumption of the ideal detector resolution. Each distribution is 
normalized to a unit area.
}
\label{fig:mmc_mmc_ideal}
\end{figure*}

To evaluate the performance of the MMC algorithm, we use inclusive $Z/\gamma^*\to\tau\tau$ and $gg\to H\to\tau\tau$ (with $M_H$=115, 120, and 130~$GeV$/c$^2$) events produced by the Pythia MC generator supplemented with the TAUOLA package. 
All events are generated for $p\bar{p}$ collisions at $\sqrt{s}=$1.96 TeV. However, the algorithm performance for events
produced in $pp$ collisions at the LHC is expected to be very similar to that for $\tau\tau$ events at the Tevatron.   
Unless it is otherwise noted, we select events where both visible $\tau$’s have $p_T$$>$10~GeV/$c$ and $\met$$>$10~GeV ($\met$ is calculated as a combined transverse momentum of all neutrinos from both $\tau$ decays). The events are categorized according to the decay mode of each of the two $\tau$ leptons (leptonic, 1-prong or 3-prong hadronic), and the $\tau\tau$ mass is reconstructed using the appropriate version of the algorithm. Results for $H\to\tau\tau$ events with $M_H=$115~GeV/c$^2$ are shown in Fig.~\ref{fig:mmc_mmc_ideal} for each of the three decay categories. In all cases, the peak position of the reconstructed $M_{\tau\tau}$ distribution is within $\sim$2$\%$ of the true mass, indicating that the assumptions used in the algorithm do not bias the reconstructed mass. The resolution of the reconstructed $M_{\tau \tau}$, defined as the RMS of the mass distribution in the $(1.0\pm0.4)\times M_{\tau\tau}^{true}$ range, changes from $\sim$8$\%$ for events with both $\tau$'s decaying hadronically to $\sim$13$\%$ when both $\tau$'s decay leptonically. The worse resolution in the leptonic modes is due to the weaker constraints on the system. The fraction of events where Eqs.~\ref{eq:math_mms} cannot be solved for any of the grid points ranges from $\sim$1$\%$ to 3$\%$, which demonstrates the high reconstruction efficiency of the MMC algorithm. Figure~\ref{fig:mmc_mmc_ideal} shows comparison of the reconstructed $\tau\tau$ mass in $Z/\gamma^{*}$ and Higgs boson events with $M_H=$115 and 130~GeV/c$^{2}$ when both $\tau$'s decay hadronically. 

\subsection{Effects of Detector Resolution}
\label{sec:MMCdetector}

To evaluate the importance of detector effects on MMC performance, we use the same inclusive $Z/\gamma^*\to\tau\tau$ and $gg\to H\to\tau\tau$ events and smear the $\met$ and momenta of the visible $\tau$ decay products according to typical detector resolutions\footnote{For simplicity, we assume Gaussian detector resolutions in this study.} at the LHC and Tevatron experiments~\cite{atlas_tdr,cms_tdr,cdf_tau,cms_tau}. We assume 3$\%$ and 10$\%$ resolutions for momenta of light leptons and hadronic $\tau$-jets, respectively. The $\met$ resolution for each of the two ($x$- and $y$-) components is taken to be $\sigma_x$=$\sigma_y$=$\sigma$=5~GeV~\cite{cdf_resolutions}. Note that in a real experimental environment, the mismeasurements 
in lepton or hadronic $\tau$-jet momenta also lead to an additional mismeasurement in $\met$. This effect is properly accounted 
for in our studies. Angular resolutions for visible $\tau$ decay products of typical detectors are usually accurate enough to 
have no noticeable effect on our calculations.

\begin{figure*}[htb]
\includegraphics[width=0.50\linewidth]{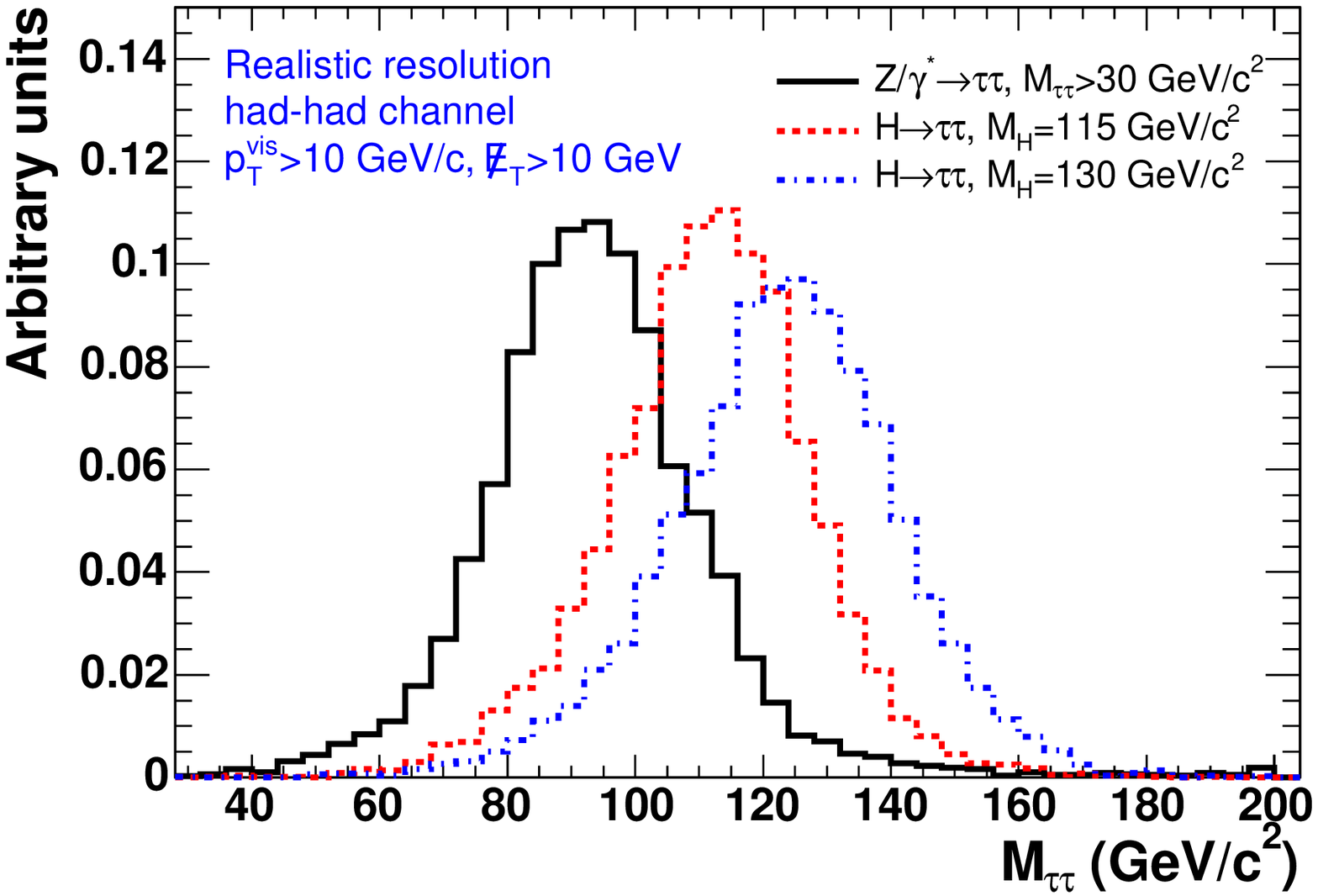}
\includegraphics[width=0.50\linewidth]{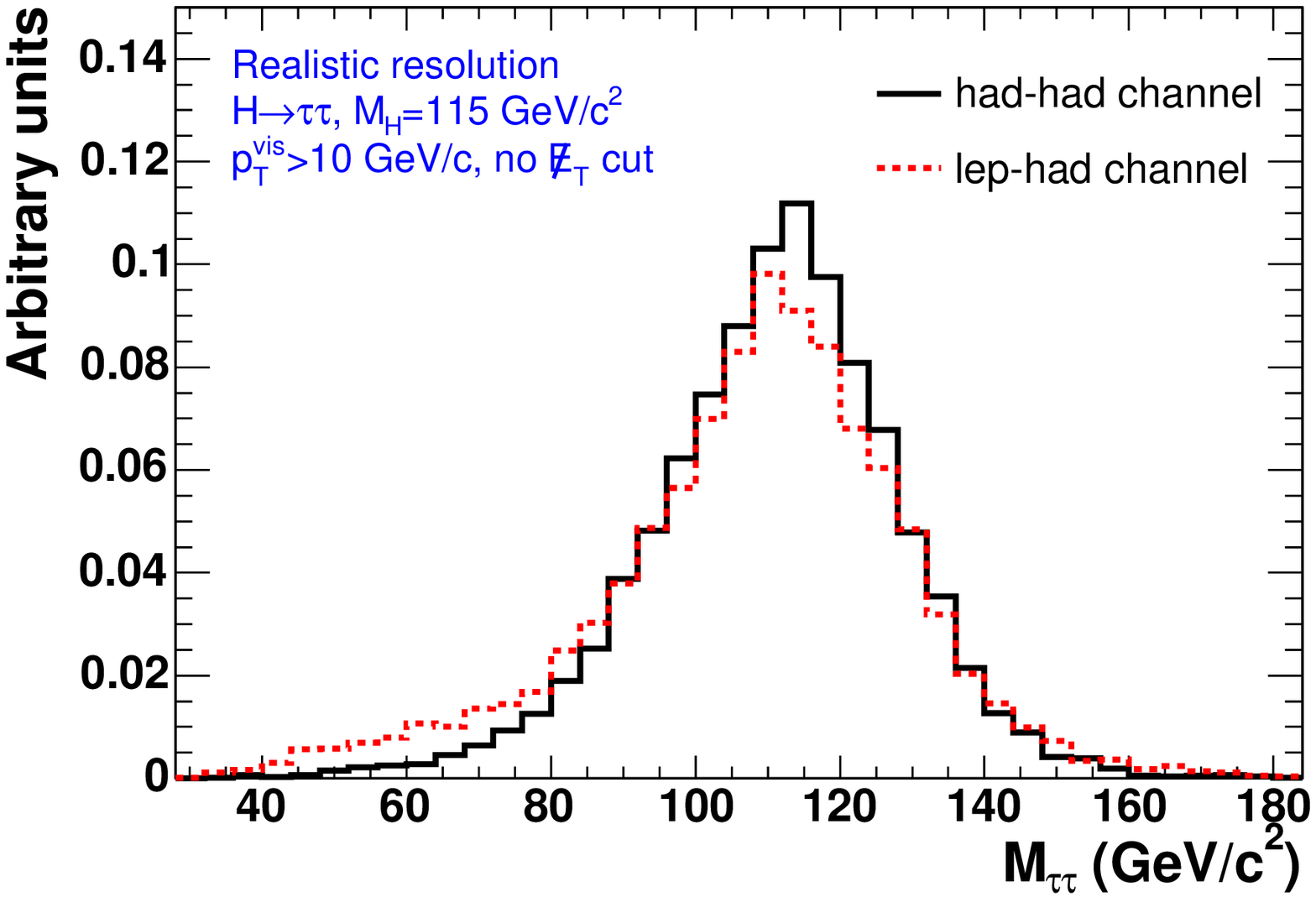}
\caption{Left plot illustrates the reconstructed $M_{\tau\tau}$ mass in $Z\to\tau\tau$ (solid line) and $H\to\tau\tau$ 
events with $M_H=$115 (dashed line) and 130~GeV/c$^{2}$ (dashed-dotted line) in the fully hadronic decay mode. Right plot 
demonstrates a comparison of the reconstructed mass in $H\to\tau\tau$ events with $M_H=$115~GeV/c$^{2}$ when both $\tau$'s 
decay hadronically (solid line) and when one $\tau$ decays leptonically and the other one hadronically (dashed
line). All results are obtained by using the MMC technique in events simulated with the realistic detector resolution. 
Each distribution is normalized to a unit area.}
\label{fig:mmc_toysim}
\end{figure*}

We find that mismeasurements of the momentum of $\tau$ lepton decay products alone have little effect on the performance of the algorithm. The $M_{\tau\tau}$ peak position and resolution are nearly unaffected and the efficiency is decreased by $\sim$3-7$\%$ as a result of mismeasurements in the momenta of visible $\tau$ decay products, which are also propagated into the $\met$. The stability of the peak position is related to a built-in self-correcting mechanism in the algorithm, which compensates slight under(over)-estimations in the measured momenta of visible decay products by over(under)-estimating the missing momentum, thus leading to the correctly reconstructed momentum of the original $\tau$ lepton. 

One could expect the effects of finite $\met$ resolution to degrade the algorithm performance. We find that, if not taken into account, a 5~GeV resolution in $\met$ results in a 30-40$\%$ drop in reconstruction efficiency, long tails in the reconstructed $M_{\tau\tau}$, and a significant degradation in the $M_{\tau\tau}$ resolution (e.g., from $\sim$8$\%$ to $\sim$18$\%$ in fully hadronic $\tau\tau$ decay mode). In particular, a large reduction in the reconstruction efficiency occurs because mismeasurements in $\met$ break the key assumption that the neutrinos from the $\tau$ decays are the sole source of $\met$ in the event (see Sec.~\ref{sec:MMCconcept} and Eqs.~\ref{eq:math_mms}). To mitigate these effects, the implementation of the MMC technique in a realistic experimental environment has to be adjusted to allow for possible mismeasurements in $\met$. It is achieved by increasing the dimensionality of the parameter space in which the scanning is performed to include the two components of the $\met$ resolution (for $\met_x$ and $\met_y$). In this case, the event likelihood, $\cal L$, has to be augmented with the corresponding resolution functions:
\begin{eqnarray}
{\cal L}=-\log{({{\cal P}(\Delta R_1,p_{\tau1})}\times{{\cal P}(\Delta R_2,p_{\tau2})}\times{{\cal P}(\Delta\met_x)}\times{{\cal P}(\Delta\met_y)})},
\end{eqnarray}
where the probability functions ${\cal P}(\Delta\met_x)$ and ${\cal P}(\Delta\met_y)$ are defined as:
\begin{eqnarray}
{\cal P}(\met_{x,y}) = \exp{\left(-\frac{(\Delta\met_{x,y})^2}{2\sigma^2}\right)}
\label{eq:gauss} 
\end{eqnarray}
where $\sigma$ is the resolution (which we take to be 5~GeV) and $\Delta\met_{x,y}$ are variations of $x$- or $y$- components of $\met$. In a real experimental setup, the $\met$ uncertainty can be larger in a particular direction, for example, if there is an energetic jet. In such cases, the uncertainty in the jet energy measurement will increase the uncertainty in $\met$ in the direction of the jet. These effects can be accounted for by suitably defining the $x$- and $y$- directions on an event-by-event basis and by choosing the appropriate $\sigma_x$ and $\sigma_y$, which will not be equal to each other in general. 

We evaluate the performance of the modified algorithm (with the $\met$ resolution scan) using $Z\to\tau\tau$ and $H\to\tau\tau$ events smeared with the realistic detector resolutions as described above. Figure~\ref{fig:mmc_toysim} shows the distribution of the reconstructed $M_{\tau\tau}$ in the fully hadronic decay mode for three samples: inclusive $Z/\gamma^{*}\to\tau\tau$ and $gg\to H\to\tau\tau$ with $M_H$=115 and 130~GeV/c$^{2}$. Right plot in the same Fig.~\ref{fig:mmc_toysim} demonstrates a comparison of the reconstructed mass in $H\to\tau\tau$ events with $M_H$=115~GeV/c$^{2}$ in the case when both $\tau$'s decay hadronically (solid line) and in the case when one $\tau$ decays leptonically and the other one hadronically (dashed line). We find that the modified MMC algorithm recovers almost all lost efficiency (to the level of 97-99$\%$) and significantly improves the relative $M_{\tau\tau}$ resolution (to the level of $\sim$14$\%$). The reconstructed mass peak position for each of the resonances is consistent with the corresponding true mass. We also observe that the mass resolution somewhat improves (at the level of 1-2$\%$) for events with higher $\met$ and/or higher $p_T$ of visible decay products. 

\subsection{Comparisons with Existing Methods}
\label{sec:MMCcomparison}

\begin{figure*}[htb]
\includegraphics[width=0.5\linewidth]{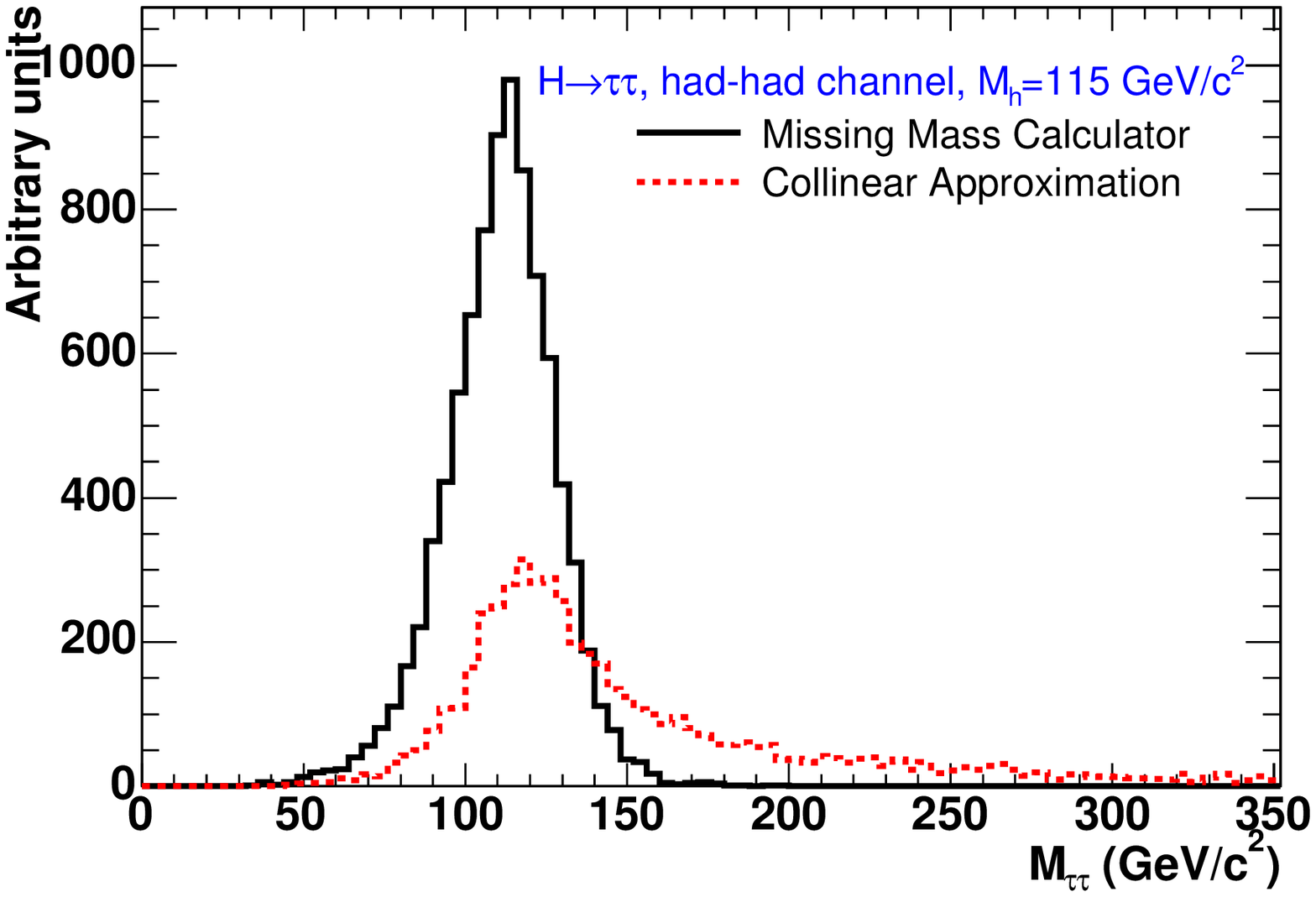}
\includegraphics[width=0.5\linewidth]{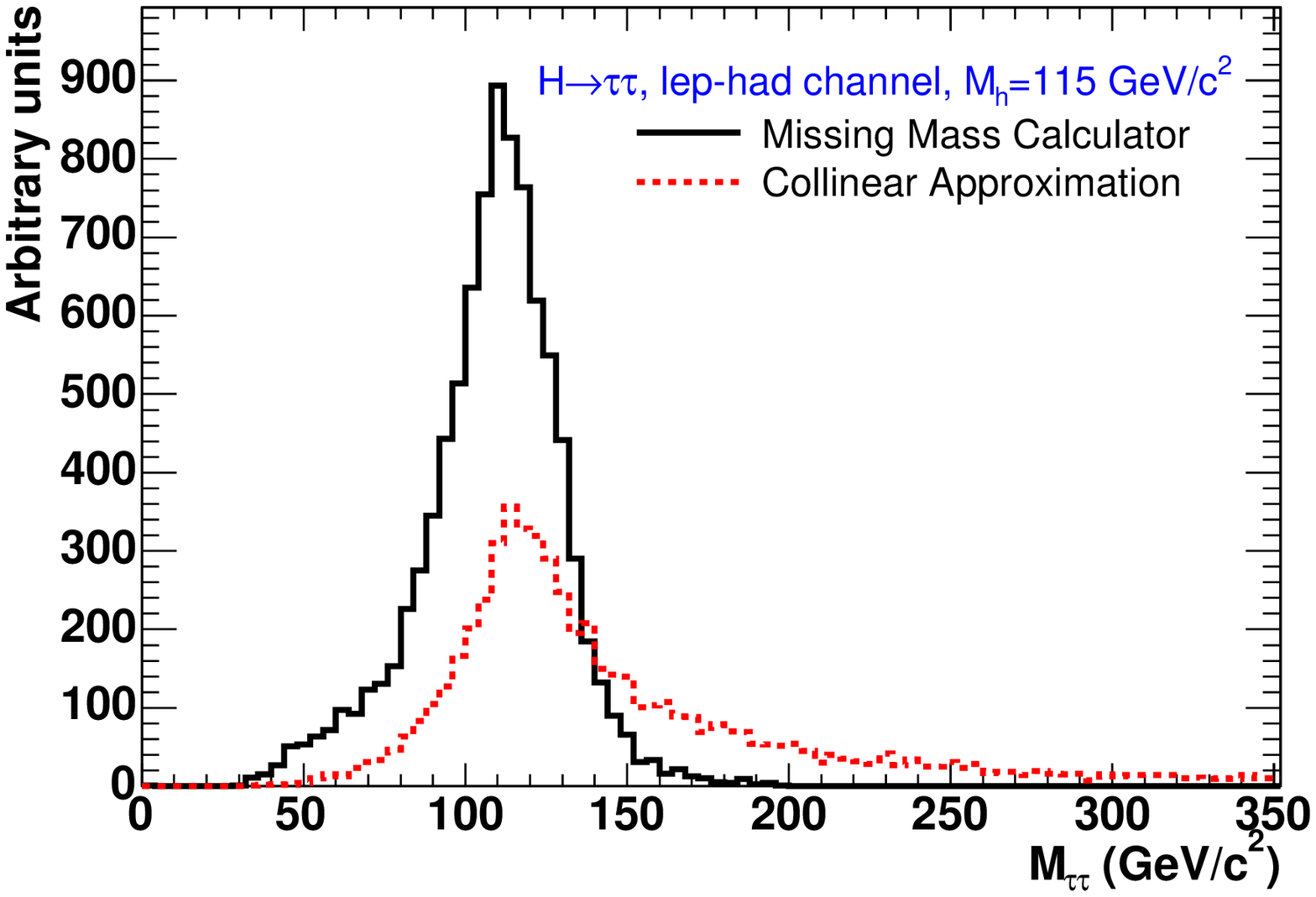}
\caption{Reconstructed mass of the $\tau\tau$ system for $gg\to H\to\tau\tau$ events with $M_H=115$ GeV/c$^2$ simulated with realistic detector resolution effects. Results of the MMC technique (solid line) are compared to those based on the collinear approximation (dashed line). Two categories of $\tau\tau$ events are considered: when both $\tau$ leptons decay hadronically (left plot), and when one of the $\tau$ leptons decays to $e$ or $\mu$ and the other $\tau$ decays hadronically (right plot). The difference in normalizations of the MMC and collinear approximation results reflects a higher efficiency of the MMC method. A long tail in the $M_{\tau\tau}$ distribution for the collinear approximation is due to the events where the two $\tau$ leptons have approximately back-to-back topology.
\label{fig:mmc_mass}}
\end{figure*}

\begin{figure*}[htb]
\includegraphics[width=0.45\linewidth]{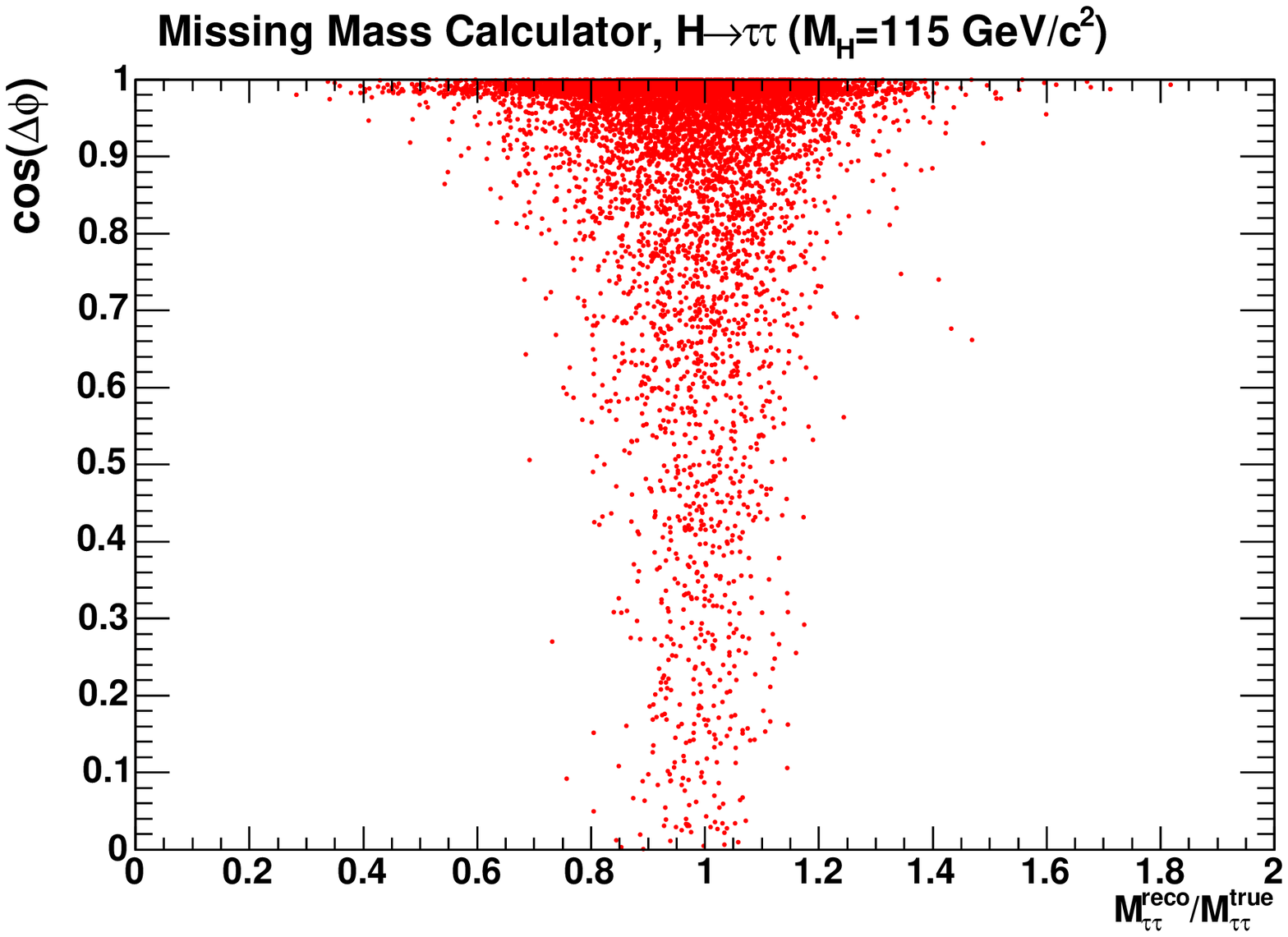}
\includegraphics[width=0.45\linewidth]{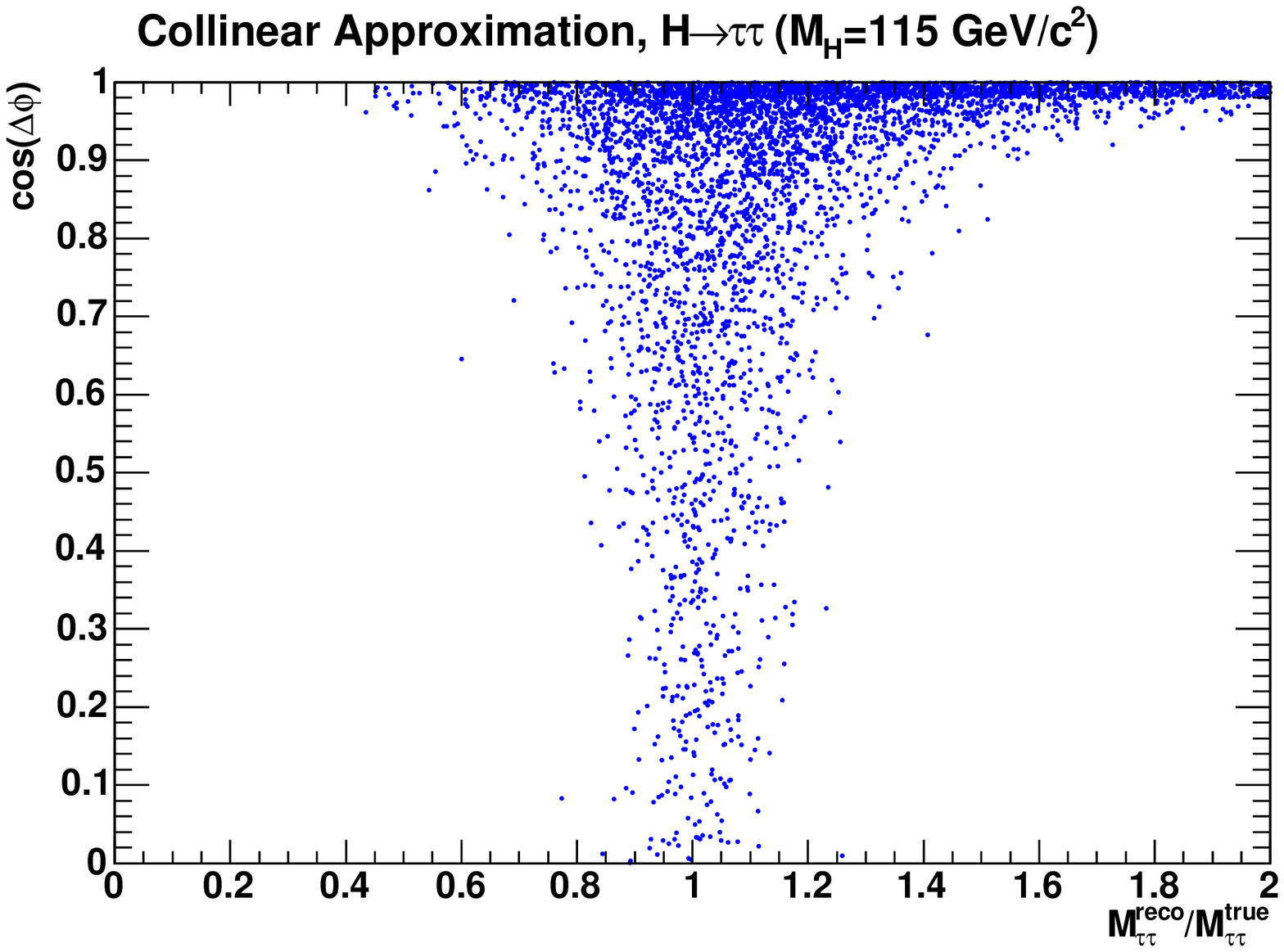}
\caption{Distribution of the ratio of the reconstructed invariant mass $M_{\tau \tau}$ versus $\cos{\Delta\phi}$, where $\Delta\phi$ is the azimuthal angle between visible decay products of the two $\tau$ leptons in $H\to\tau\tau$ events with $M_h$=115~GeV/c$^2$. Results of the MMC method (left plot) are compared to those of the collinear approximation (right plot). 
Note that the new method performs significantly better for nearly back-to-back topology ($\cos{\Delta\phi}\to$1), which 
constitutes the bulk of all $\tau\tau$ events.
\label{fig:mmc_m_vs_sin_theta}}
\end{figure*}

Figure~\ref{fig:mmc_mass} shows the reconstructed $M_{\tau\tau}$ distributions in $H\to\tau\tau$ events with $M_h$=115~GeV/c$^2$ obtained by using the MMC algorithm (black histogram) and the collinear approximation (red line). The events are simulated with the realistic detector resolution effects as described in the previous section. Two categories of $\tau\tau$ events are considered: both $\tau$ leptons decay hadronically (left plot in Fig.~\ref{fig:mmc_mass}), and one leptonic and one hadronic $\tau$ decay (right plot in Fig.~\ref{fig:mmc_mass}). The difference in normalizations of the MMC and collinear approximation results reflects a higher efficiency (by a factor of $\sim$1.7) of the MMC method. This is because the substantial fraction of events have a moderate $\met$ or approximately back-to-back topology and are non-reconstructible by the collinear approximation technique. This happens when small mismeasurements in $\met$ lead to configurations for which Eqs.~\ref{eq:math_collinear_approximation} have no solution. In contrast, the MMC method resolves this problem and has an average efficiency of 97-99$\%$. In addition to a better resolution in the core of the $M_{\tau\tau}$ distribution, an important feature of the MMC technique is the absence of the long tail toward higher masses present in the distribution obtained using the collinear approximation. This tail is associated with the events of approximately back-to-back topology, where the collinear approximation diverges as $\cos{\Delta\phi}\to$1. ($\Delta\phi$ is the angle between two visible $\tau$ decay products in the plane transverse to the beam line.) The reason for this divergency is discussed in Sec.~\ref{sec:CollApp}. The effect is illustrated in Fig.~\ref{fig:mmc_m_vs_sin_theta}, which shows a comparison of the ratio of the reconstructed and true mass as a function of $\cos{\Delta\phi}$ for the two methods. In contrast to the collinear approximation, the absence of long tails toward large masses in the MMC technique presents a significant improvement 
for low-mass Higgs boson searches in the $H\to\tau\tau$ channel by significantly reducing a large $Z\to\tau\tau$ background, which would otherwise completely overwhelm the Higgs search region.

It is also important to point out that the algorithm efficiency and the shapes of likelihood ${\cal L}$ distributions are expected to be different for events with true $\tau$ leptons and those where jets are misidentified as hadronically decaying $\tau$ leptons. This may offer an additional handle on the backgrounds with the misidentified $\tau$ leptons, most notably $W+$jets and QCD multi-jet events, and it needs to be further investigated. 

\section{Performance With Data and Monte Carlo After Full Detector Simulation}
\label{sec:Data}

To illustrate the power of the proposed method using real data, we select a sample of clean $Z/\gamma^{*} \to  \tau \tau$ events collected by the CDF experiment~\cite{cdf} in $p\bar{p}$ collisions at a center-of-mass energy $\sqrt{s}$=1.96~TeV at the Tevatron. We obtain a high purity sample of $\tau\tau$ events in the channel where one of the $\tau$ leptons decays into a light lepton ($e$ or $\mu$) while the other decays into one of the hadronic modes. The requirement of a well isolated muon or electron significantly reduces QCD multi-jet backgrounds in this channel. We then compare the observed $\tau\tau$ invariant mass spectrum of $Z/\gamma^{*} \to \tau \tau$ events reconstructed using the MMC technique with results obtained using the collinear approximation. Data are also compared with predictions obtained from Monte Carlo (MC) simulation. Signal events and backgrounds coming from $Z/\gamma^{*}\to ee/\mu\mu$ and $W+jets$ processes are generated by PYTHIA~\cite{pythia} Tune A with CTEQ5L parton distribution functions~\cite{cteq}. The detector response is simulated with the GEANT-3 package~\cite{geant3}. QCD multi-jet background is estimated from data by using events with lepton candidates of the same charge.

\subsection{Data Selections}
\label{sec:EvntSelection}

Full description of the CDF detector is available elsewhere~\cite{cdf}. Here we only briefly discuss the algorithms and selection specific to this study. Our choice of selection requirements is based on the two following considerations. First, backgrounds in the data sample of $Z/\gamma^{*}\to\tau\tau$ candidate events should be very low to allow unambiguous demonstration of the power of the MMC technique. Second, event selection should not be sensitive to potential Higgs boson signal to avoid biases in the currently ongoing $H\to\tau\tau$ analysis. To achieve these goals, selection requirements are chosen to be extremely tight, thus effective for only a small fraction of $Z/\gamma^{*}\to\tau\tau$ and $H\to\tau\tau$ events. In fact, the signal acceptance is reduced by a factor of 
$\sim$6 compared to that in the ongoing search for $H\to\tau\tau$. 

We use data collected with the inclusive electron and muon triggers~\cite{cdf_emu}. These select high-$p_T$ electron and muon candidates with $|\eta|$$\le$1. In the offline event selection, we require at least one reconstructed electron or muon candidate with transverse energy (for $e$'s) or momentum (for $\mu$'s) satisfying $E_T$$>$20~GeV 
or $p_T$$>$20~GeV/$c$, respectively; this ensures we are on the trigger efficiency plateau. Data used in this study correspond to an integrated luminosity of 5.6~fb$^{-1}$.

Hadronically decaying $\tau$ leptons are identified as narrow calorimeter clusters associated with one or three charged tracks. We refer to publication~\cite{cdf_tau} for details of $\tau$ identification; the selection requirements in this study differ as follows. We apply track isolation ($I_{trk}^{\Delta R}$) defined as 
a scalar sum of the transverse momenta of all tracks in a certain range of $\Delta R = \sqrt{\Delta \phi^2 + \Delta \eta^2}$ around the highest $p_T$ (seed) track of the $\tau$ candidate. The seed track is required to have 
$p_T$$>$10~GeV/$c$ and $I_{trk}^{0.17 < \Delta R < 0.52}$$<$1~GeV/$c$. To suppress events with electrons or muons misidentified as hadronic $\tau$ decays, we require $E^{EM}/(E^{EM}+E^{HAD})$$<$0.9 or 
$(E^{EM}+E^{HAD})/\sum_{trks}p$$>$0.5 respectively, where $E^{EM}$ and $E^{HAD}$ are electromagnetic and hadronic
energy deposits in calorimeter clusters associated with $\tau$ candidates.

For electron identification we apply the same selection criteria as in Ref.~\cite{cdf_tau} with the following
exceptions. We do not apply calorimeter isolation and require the electron to have $I_{trk}^{\Delta R < 0.52}$$<$1~GeV/$c$ if the hadronic $\tau$ in the event has one track or $I_{trk}^{\Delta R < 0.7}$$=$0~GeV/$c$ 
if the hadronic $\tau$ has three tracks. In addition, we require hits in the silicon vertex detector associated 
with the electron track in the case of 3-prong $\tau$ decays. For events with 1-prong $\tau$ decays, the energy 
of the calorimeter cluster associated with the electron should be consistent with track momentum such that $E^{EM}/p^{trk}$$<$1.1. Muon identification follows the procedure described in Ref.~\cite{cdf_emu}. We apply the 
track isolation, $I_{trk}^{\Delta R < 0.52}$$<$1~GeV/$c$, to all muons regardless the hadronic $\tau$ decay mode.

Additional selection criteria are applied on event kinematics to suppress backgrounds with fake electron, muon 
or $\tau$ candidates. The lepton (electron or muon) and hadronic $\tau$ are required to have opposite charges 
to reduce the QCD multi-jet backgrounds. We remove events where the electron candidate is identified as a 
conversion. We also reject events consistent with those coming from cosmic muons. To suppress $W+jets$ background, 
we require the lepton ($e$ or $\mu$) and hadronic $\tau$ to be back-to-back in the $x-y$ plane: $\Delta\phi(\mu/e-trk,\tau -trk)$$>$2.9~radians. In addition we reject events which have at least one jet with $E_T>$10~GeV and $|\eta|$$<$3.6. Jets are reconstructed in the calorimeter using the JETCLU cone algorithm~\cite{jetclu} with a cone radius of 0.4 in the ($\eta$,$\phi$) space. To reduce $Z/\gamma^{*}\to ee/\mu\mu$ background, we only select events with electron transverse energy 20~GeV$<$$E_{T}^{e}$$<$40~GeV or muon transverse momentum 20~GeV/$c$$<$$p_{T}^{\mu}$$<$40~GeV/$c$. Limiting the maximum value of the lepton transverse energy or momentum also helps to reduce acceptance of potential $H\to\tau\tau$ signal. We further suppress $Z/\gamma^{*}\to ee/\mu\mu$ background by looking at the invariant mass of every pair of tracks and every pair of clusters in the electromagnetic calorimeter. We reject an event if at least one pair has invariant mass consistent with the mass 
of a Z boson: {\it i.e.}, it is the range 66~GeV/$c^2$$<$$M$$<$111~GeV/$c^2$ in the case of tracks and 76~GeV/$c^2$$<$$M$$<$106~GeV/$c^2$ in the case of calorimeter clusters.

\begin{figure*}[htb]
\includegraphics[width=0.5\linewidth]{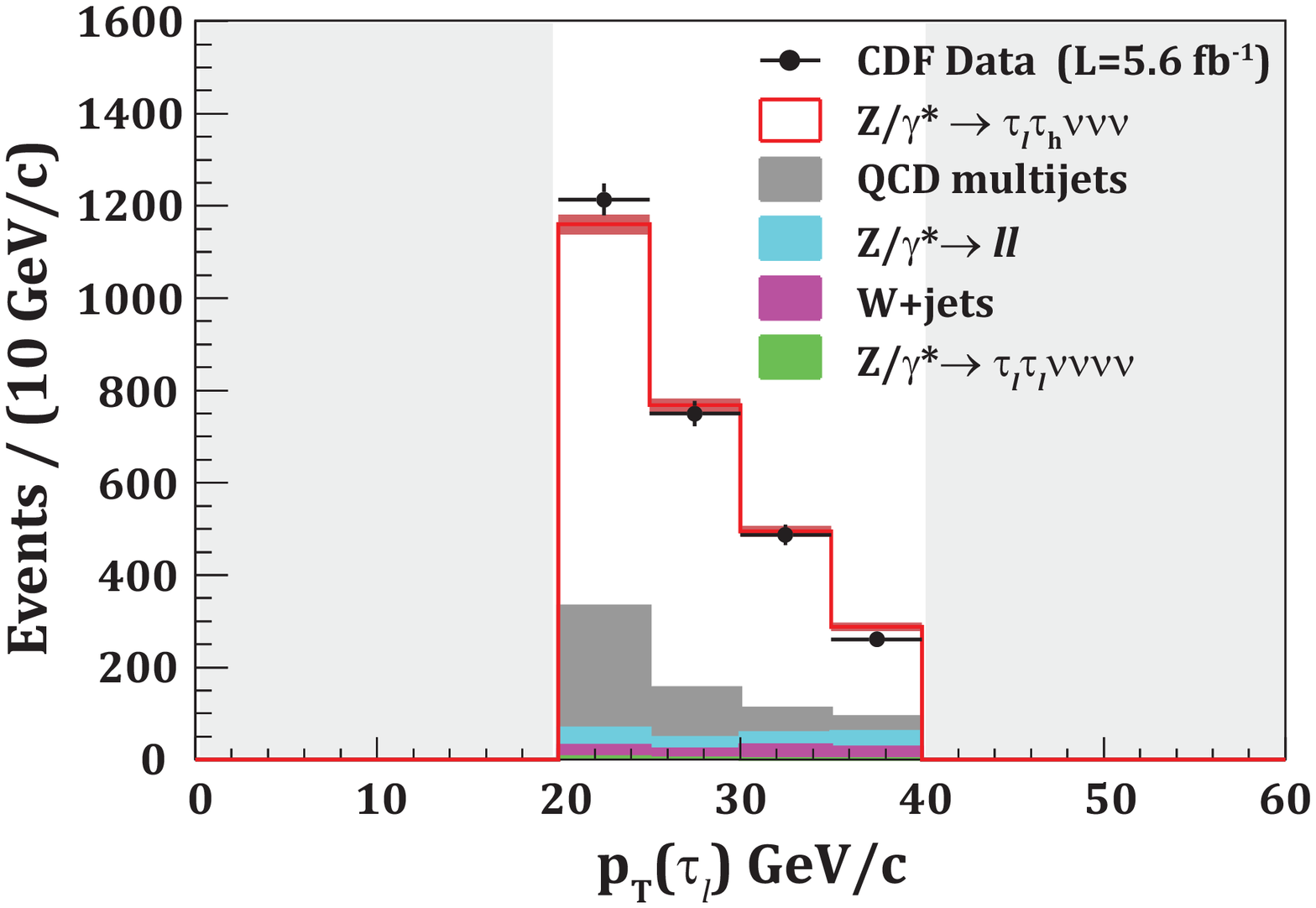}
\includegraphics[width=0.5\linewidth]{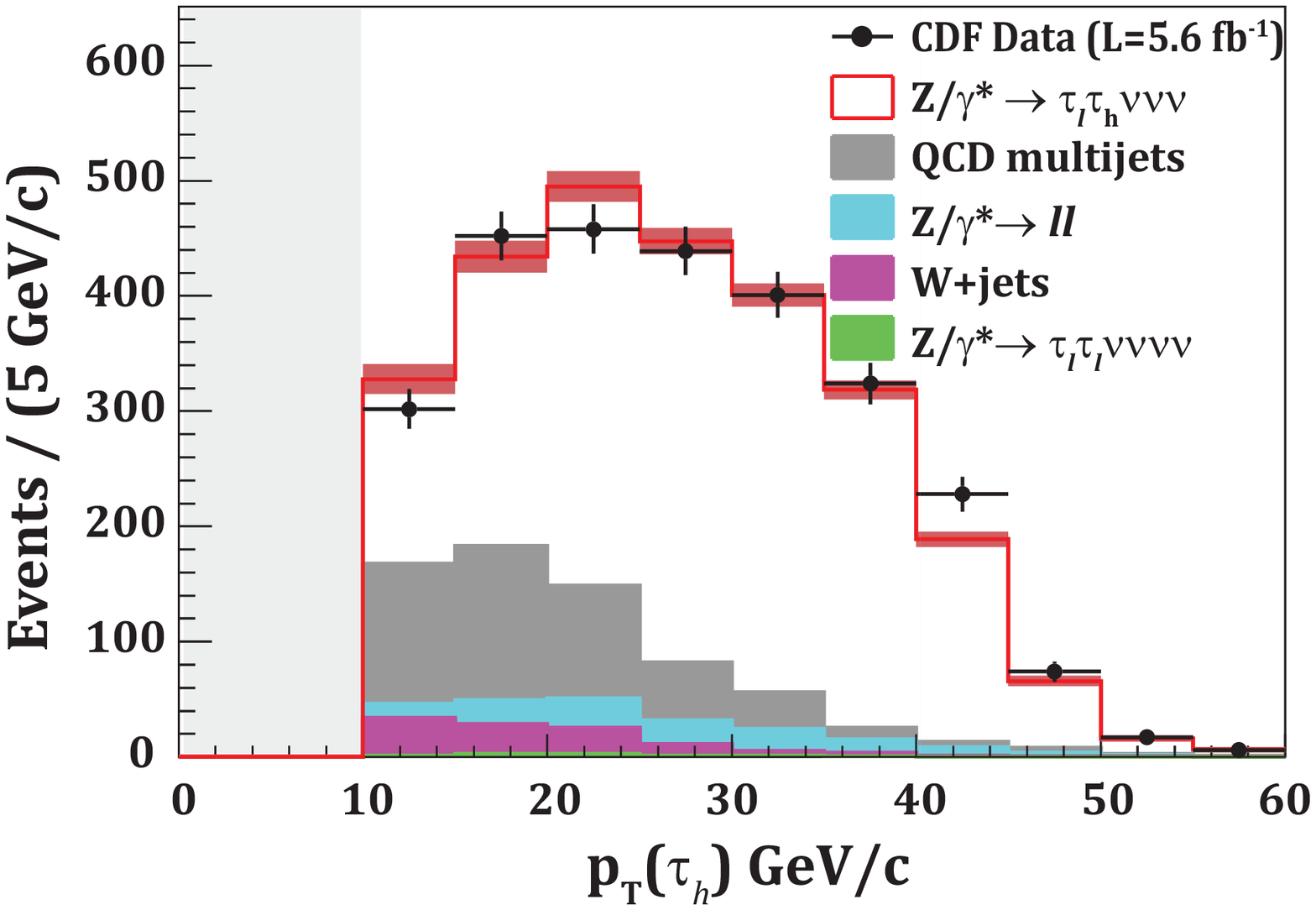}
\includegraphics[width=0.5\linewidth]{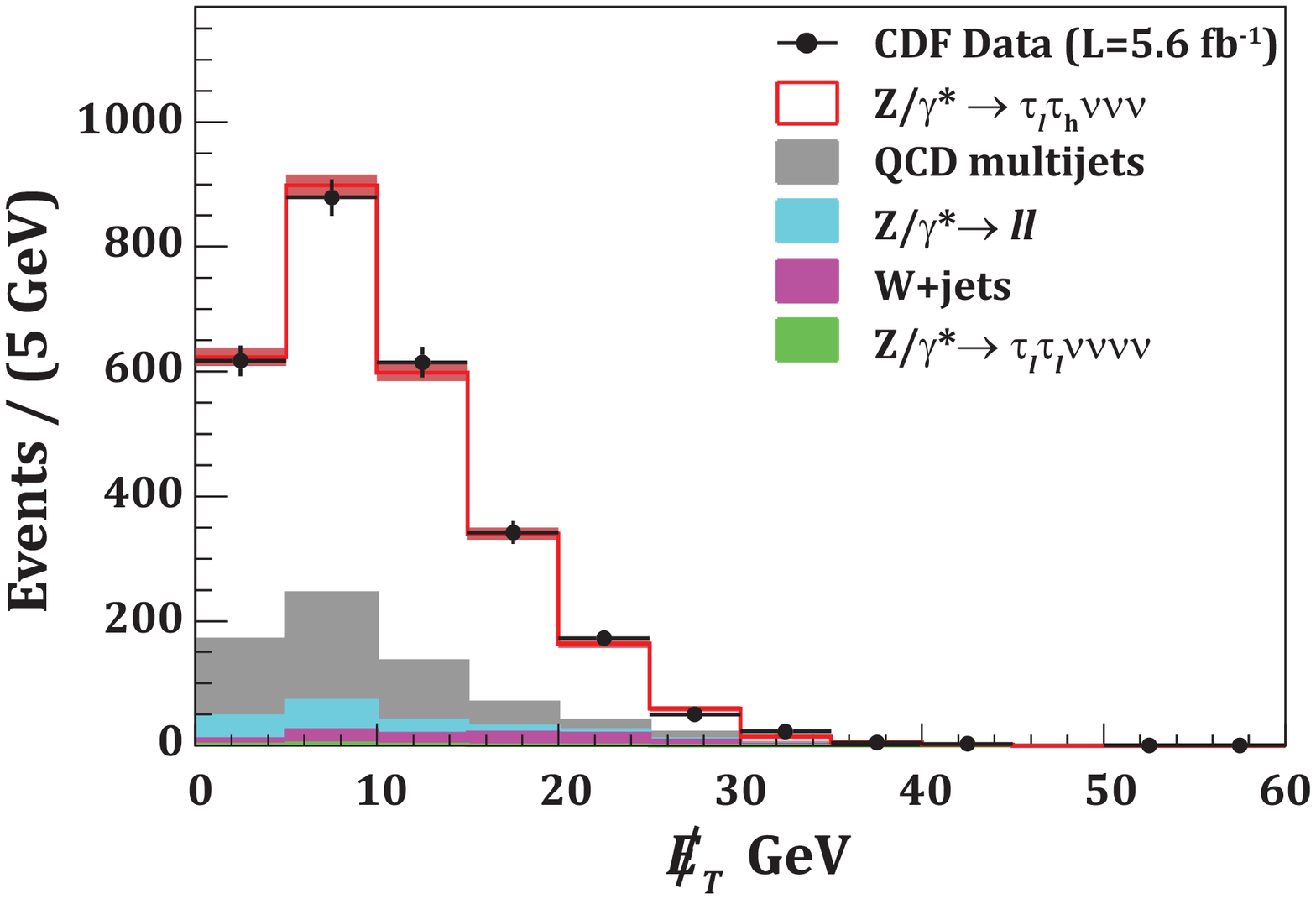}
\caption{Kinematic distributions for $Z/\gamma^{*}\to\tau\tau\to l\tau_h\nu\nu^{`}\bar{\nu}$ ($l=e$ or $\mu$) events selected from the CDF data: (a) transverse momentum of the light lepton; (b) transverse
momentum of visible decay products of the hadronically decaying $\tau$ lepton, $\tau_h$;
(c) missing transverse energy, $\met$, in the event~\cite{met}.
\label{fig:data_kinematics}}
\end{figure*}

Figure~\ref{fig:data_kinematics} shows kinematic distributions for the selected $Z/\gamma^{*}\to\tau\tau$ candidate events demonstrating the high purity of the sample. The MC predictions for signal and background events are normalized
based on the product of integrated luminosity and corresponding production cross sections. 

\subsection{Mass Reconstruction using the MMC Technique}
\label{sec:MMCdata}

The $\tau\tau$ invariant mass in data and MC events with full detector simulation is reconstructed with the MMC
algorithm described in Sec.~\ref{sec:MMCdetector}. Although $Z/\gamma^{*}\to\tau\tau$ events in our data sample 
have no jets, a pair of $\tau$ leptons may be accompanied by one or more jets when different event selection 
requirements are applied. Therefore, we describe the $\met$ resolution parametrization for events with $N_{jet}$=0 
and $N_{jet}$$>$0. For this purpose, we only count jets with $E_T$$>$15~GeV and $|\eta|$$<$3.6.

For $N_{jet}$=0 events, we perform scans for the $x$- and $y$-components of $\met$. The corresponding resolutions of
each $\met$ component are parametrized by Gaussian distributions (Eq.~\ref{eq:gauss}) with width $\sigma_{UE}$, which is a function of unclustered energy\footnote{The unclustered energy is defined as the scalar sum of $E_T$ for all calorimeter towers which are not included in electron, jet or hadronic $\tau$ reconstruction.} in the event: $\sigma_{UE}$=$p_0+p_1\sqrt{\sum E_T}$. We use the same values of $p_0$ and $p_1$ as reported in the CDF publication~\cite{cdf_resolutions}.

In events with $N_{jet}$$>$0, we consider the $\met$ resolution in the directions parallel ($\sigma_{\parallel}$) and perpendicular ($\sigma_{\perp}$) to the direction of a leading jet in the event. We take $\sigma_{\perp}$=$\sigma_{UE}$ and $\sigma_{\parallel}$=$\sqrt{\sigma_{UE}^2+\sigma_{jet}^2}$, where $\sigma_{jet}$ is the jet energy resolution which is a function of the jet $E_T$ and $\eta$. For $\sigma_{jet}$, we use a simplified version (assuming Gaussian jet energy resolution) of the parametrization reported in Ref.~\cite{cdf_resolutions}. 
If there is more than one jet in the event, we project $\sigma_{jet}$ for each additional jet onto axes 
parallel and perpendicular to the leading jet direction. These projections are then added in quadrature to $\sigma_{\parallel}$ and $\sigma_{\perp}$, respectively. Finally, we perform scans for $\met$ components parallel 
and perpendicular to the leading jet direction.      

\subsection{Reconstructed Mass Spectrum in Data}
\label{sec:MMCperformance}

\begin{figure*}[htb]
\includegraphics[width=0.5\linewidth]{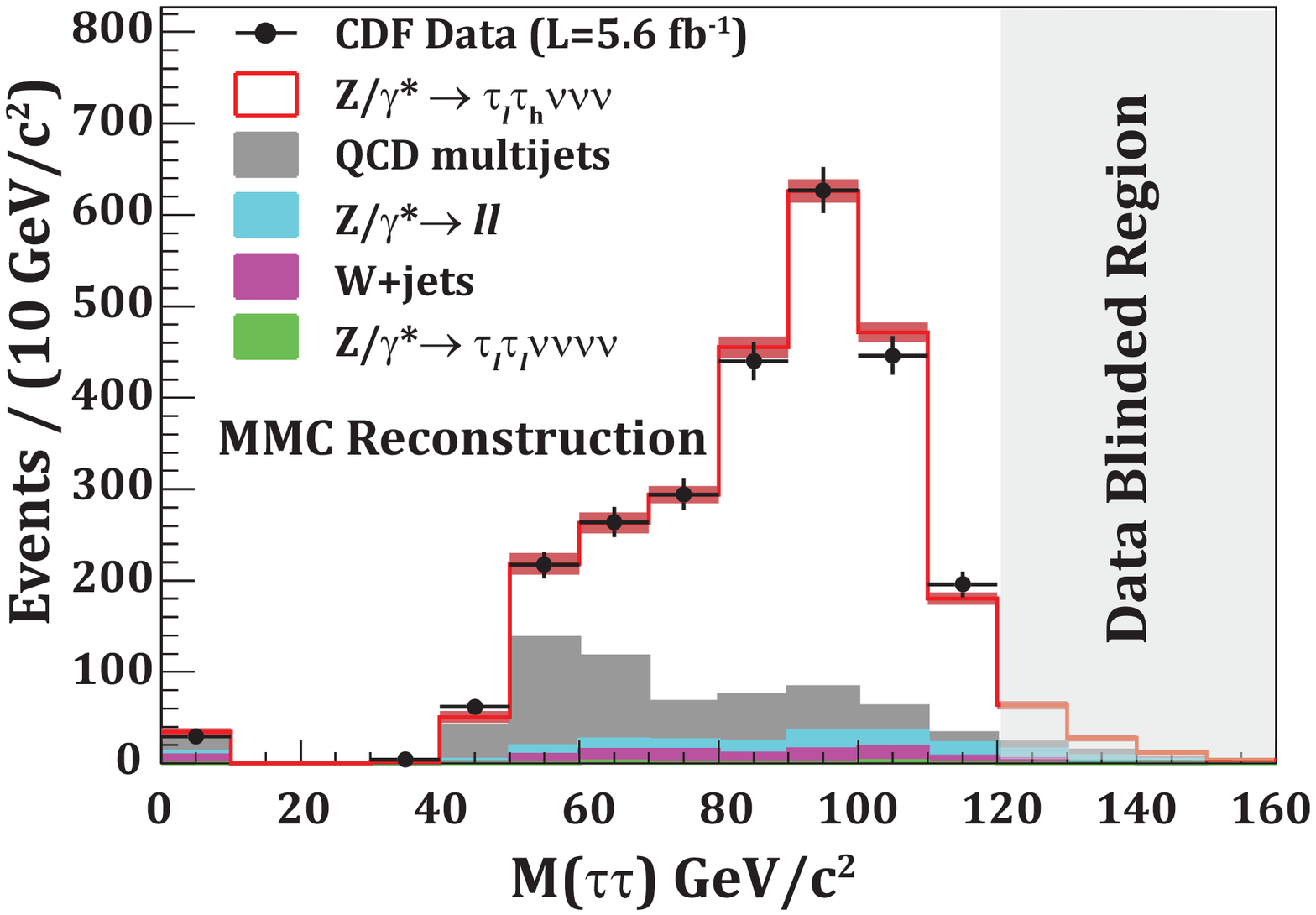}
\includegraphics[width=0.5\linewidth]{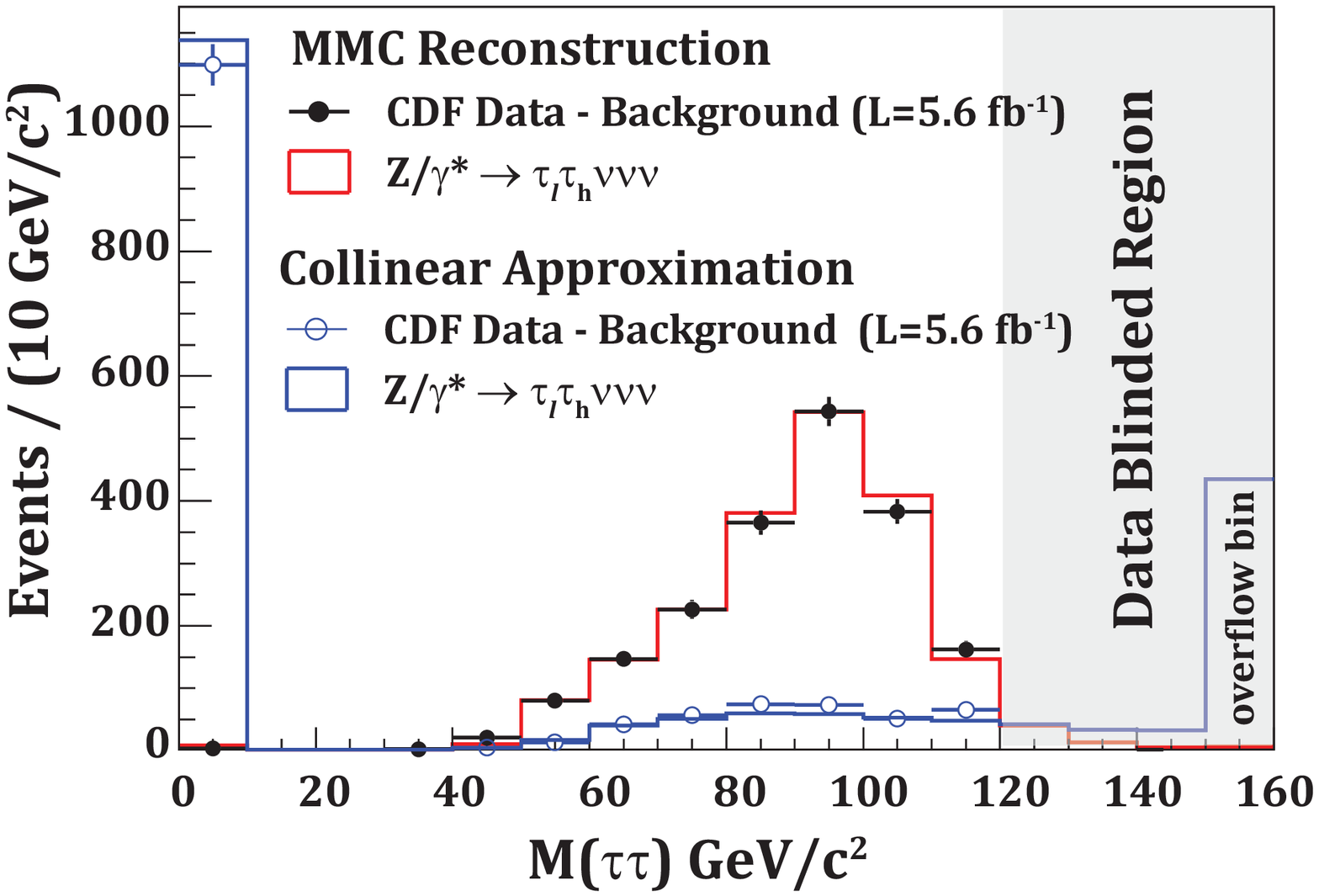}
\caption{Reconstructed mass of the $\tau\tau$ system in $Z/\gamma^{*}\to\tau\tau\to l\tau_h\nu\nu^{`}\bar{\nu}$ 
($l=e$ or $\mu$) candidate events using the MMC and collinear approximation techniques: (a) $\tau\tau$ mass reconstructed with MMC technique, data (points) compared to the sum of background and signal predictions; (b) comparison of the MMC (filled circles are data and red line is the signal prediction) and collinear approximation (open circles are data and blue line is the signal prediction) results after subtracting the corresponding 
background predictions. Unreconstructed events are shown in the first histogram bin ($M_{\tau\tau}$$\sim$0).
Events with $M_{\tau\tau}$$>$160~GeV/$c^{2}$ are outside the histogram range and are shown in the overflow bin.
\label{fig:data_mass}}
\end{figure*}

Figure~\ref{fig:data_mass} shows the $\tau\tau$ invariant mass distribution obtained with the MMC and collinear
approximation methods for our data sample of $Z/\gamma^{*}\to\tau\tau$ events. The left plot shows $\tau\tau$ 
mass calculated with the MMC technique and compares data with the sum of background and signal predictions. The 
first bin of the distribution contains events where no solution for $M_{\tau \tau}$ was found. We note that events unreconstructed by the MMC method are predominantly from background processes. 

Excellent performance of the MMC technique and its advantage over the collinear approximation in terms of 
resolution and reconstruction efficiency is clearly demonstrated by differences in shape and normalization
of the $M_{\tau\tau}$ distributions in the right plot of Fig.~\ref{fig:data_mass}. To facilitate a comparison, 
the background predictions are subtracted from the $M_{\tau\tau}$ distributions in data. Events with the 
reconstructed mass $M_{\tau\tau}$$>$160~GeV/$c^{2}$ are outside the histogram range and are shown in the overflow 
bin. The fraction of such events is negligible ($\sim$0.3$\%$) for the MMC method, while it is $\sim$18$\%$ fot 
the collinear approximation. Shapes of the distributions agree well between data and simulation, therefore we use simulated $Z/\gamma^{*}\to\tau\tau$ events to estimate the resolution and efficiency achieved by the MMC technique. 
We follow the definition introduced in Sec.~\ref{sec:MMC} and define resolution as the RMS of the $M_{\tau\tau}/M_{\tau\tau}^{true}$ and distribution in the 0.6-1.4 range, where $M_{\tau\tau}$ is the reconstructed mass and $M_{\tau\tau}^{true}$ is the generated mass. We find the resolution to be $\sim$16$\%$ and the reconstruction efficiency to be $\sim$99$\%$, in good agreement with the results obtained using the simplified detector simulation model in Sec.~\ref{sec:MMCdetector}. In contrast, the reconstruction efficiency of the collinear approximation method is found to be $\sim$42$\%$. As explained in Sec.~\ref{sec:MMCdetector}, events where the two $\tau$'s are back-to-back in the $x-y$ plane are particularly challenging for the collinear approximation; however, such 
events represent a major fraction of the $H\to\tau\tau$ signal. A reliable $\tau\tau$ mass reconstruction with the collinear approximation is possible only for a small fraction of boosted $\tau\tau$ events (with smaller angles between the $\tau$'s, $\Delta\phi(\tau\tau)$, and higher values of $\met$). The MMC method does not have such limitations, thus giving a substantial increase in the signal acceptance. 

\section{Conclusions}
\label{sec:Conclusions}

The Missing Mass Calculation method is a novel experimental technique
proposed for reconstructing the invariant mass of resonances decaying
to a pair of $\tau$ leptons. The new method provides a substantially
more accurate reconstruction of the mass of the $\tau\tau$
system compared to commonly used techniques. Its applicability to
nearly all possible final state topologies without loss in resolution
significantly improves experimental acceptance for future searches for
resonances decaying to a pair of $\tau$ leptons. The new method
eliminates the long tail towards higher masses present in the
frequently used collinear approximation technique, thus promising a 
significant improvement in the sensitivity of $H\to\tau\tau$ searches 
at the LHC and the Tevatron, where the main challenge is the promotion 
of $Z\to\tau\tau$ background events into the Higgs boson mass region.

\section*{Acknowledgments}
We would like to thank Andre~M.~Bach and Mark~J.~Tibbetts for reading this manuscript and
providing valuable comments. We also thank the staffs of the Fermi 
National Accelerator Laboratory and Ernest Orlando Lawrence Berkeley 
National Laboratory, where a part of the work on the paper was
performed. This work would not be possible without the support of the 
U.S. Department of Energy.
%
%

\def\Journal#1#2#3#4{{#1} {\bf #2}, #3 (#4)}
\def\NCA{Nuovo Cimento}
\def\NIM{Nucl. Instrum. Methods}
\def\NIMA{{Nucl. Instrum. Methods} A}
\def\NP{Nucl. Phys.} 
\def\NPB{{Nucl. Phys.} B}
\def\PLB{{Phys. Lett.}  B}
\def\PRL{Phys. Rev. Lett.}
\def\RPP{Rep. Prog. Phys.}
\def\PRD{{Phys. Rev.} D}
\def\PR{Phys. Rep.}
\def\PRP{Prog. Theor. Phys.}
\def\ZPC{{Z. Phys.} C}
\def\MPL{{Mod. Phys. Lett.} A}
\def\EPJC{{Eur. Phys. J.} C}
\def\CPC{Comput. Phys. Commun.}

\renewcommand{\baselinestretch}{1}


\begin{thebibliography}{99}
\bibitem{met} In data and MC events with full detector simulation, the missing transverse 
energy, $\met$, is defined as an energy imbalance in the calorimeter and it is a signature 
of neutrinos that do not interact with the detector material. The $\met$ is calculated from 
all calorimeter towers with $E_T$$>$0.1~GeV in the region $|\eta|$$<$3.6 according to 
$\vec{\met}=-\sum_{i}{E}_T^i{\vec{n}_i}$ where ${\vec{n}_i}$ is a unit vector that points 
from the interaction vertex to the $i^{th}$ calorimeter tower in the transverse plane. To 
improve resolution and reduce the number of events with large fake $\met$, we apply 
corrections to the $\met$ to account for a non-linearity of the detector response for 
hadronic $\tau$'s and jets with $E_T$$>$15~GeV and for presence of reconstructed muons, 
which do not deposit their total energy in the calorimeter.
\end{thebibliography}
\end{document}